# Graphene oxide two dimensional films for thermo-optic photonic integrated devices

*David J. Moss*

Optical Sciences Centre, Swinburne University of Technology, Hawthorn, VIC 3122, Australia

Email: dmoss@swin.edu.au

**Abstract**

Efficient heat management and control in optical devices, facilitated by advanced thermo-optic materials, is critical for many applications such as photovoltaics, thermal emitters, mode-locked lasers, and optical switches. Here, we investigate the thermo-optic properties of 2D graphene oxide (GO) films by precisely integrating them onto microring resonators (MRRs) with control over the film thicknesses and lengths. We characterize the refractive index, extinction coefficient, thermo-optic coefficient, and thermal conductivity of for the GO films with different layer numbers and degrees of reduction, including reversible reduction and enhanced optical bistability induced by photo-thermal effects. Experimental results show that the thermo-optic properties of 2D GO films vary widely with the degree of reduction, with significant polarization anisotropy, enabling efficient polarization sensitive devices. The versatile thermo-optic response of 2D GO substantially expands the scope of functionalities and devices that can be engineered, making it promising for a diverse range of thermo-optic applications.

## 1. Introduction

The management and control of heat [1] is important for tackling issues like global warming, energy shortages, and device overheating. Thermo-optic effects have underpinned the functionality and efficiency of many optical devices and systems such as photovoltaics [2, 3], thermal emitters [4, 5], mode-locked lasers [6, 7], optical switches [8, 9], logic gates [10, 11], power limiters [12, 13], optical memories [14, 15], and sensors [16, 17].

Over the past two decades, the rapid advances in material science have provided powerful tools to efficiently manipulate heat transfer in optical devices [18, 19]. This has been particularly facilitated by the emergence of many novel 2D materials with atomically thin film thicknesses, such as graphene, transition metal dichalcogenides (TMDCs), and hexagonal boron nitride (hBN) [20, 21]. These 2D materials exhibit many remarkable thermo-optic properties, including the ability to tune the refractive index, absorption, conductivity and anisotropy. These have enabled a range of high performance devices with new functionality compared to conventional bulk materials [22, 23].

As a common derivative of graphene, graphene oxide (GO) is a rising star in the 2D material family [24-26]. It exhibits many distinctive optical properties, such as a broadband response, high optical nonlinearity, and significant anisotropy [27-29]. There is also a high degree of flexibility in engineering GO's properties through reduction and doping processes [17, 30, 31], offering a flexible platform for engineering a wide range of functionalized materials. Moreover, facile solution-based synthesis processes and transfer-free film coating with precise control have been developed for 2D GO films [32-34], with a high degree of compatibility with integrated device platforms.

In this paper, we comprehensively investigate a range of thermo-optic properties of 2D GO films by integrating them onto silicon nitride (SiN) microring resonators (MRRs) with precise control over the film thicknesses and lengths. We characterize the refractive index, extinction coefficient, thermo-optic coefficient, and thermal conductivity of 2D layered GO films for different layer numbers and degrees of reduction. Experimental results show that the properties, including the fundamental thermo-optic response itself, of 2D GO films vary widely with temperature as the reduction degree increases, including an increase of ~0.2280 in the refractive index, a more than 36-fold increase in the extinction coefficient, a transition from a positive thermo-optic coefficient to a negative one, and a ~48 times improvement in the thermal conductivity. In addition, the GO films shows a markedly anisotropic response for light in transverse electric (TE) and transverse magnetic (TM) polarizations, including a difference of ~0.14 for the refractive index, a ratio of ~4 for the extinction ratio, a ratio of ~6 for the thermo-optic coefficient, and a ratio of ~18 for the thermal conductivity. We also characterize reversible GO reduction and optical bistability in the hybrid MRRs induced by photo-thermal effects. Experimental findings reveal that reversible GO reduction can be initiated by light power within a specific range, and that the hybrid MRRs exhibit significantly enhanced optical bistability compared to uncoated MRRs. These results offer interesting insights for the versatile thermo-optic properties of 2D layered GO films, which offer signicant potential for thermo-optic applications.

## 2. Device design and fabrication

**Figure 1a** illustrates the atomic structures and bandgaps of GO, semi-reduced GO (srGO), and totally reduced GO (trGO). As a derivative of graphene, GO contains various oxygen functional groups (OFGs), such as hydroxyl, epoxide, carbonyl, and carboxylic groups, located either on the carbon basal plane or at the sheet edges [24]. Unlike graphene, which possesses a zero bandgap, the presence of the OFGs in GO results in a large optical bandgap $\Delta E$ typically ranging between ~2.1 eV and ~3.6 eV [26, 35]. Due to its large optical bandgap, GO exhibits low linear optical absorption at infrared wavelengths, which is about 2 orders of magnitude lower than that of graphene [36]. The reduction of GO leads to the dissociation of OFGs, along with the decreasing of $\Delta E$ and the changing of the material properties such as refractive index, optical absorption, and thermal conductivity [25]. In practical, the reduction of GO films can be realized by using various thermal reduction, chemical reduction, or photoreduction methods [24]. The bandgap and optical properties of trGO are similar to graphene, with slight differences arising from the defects within the carbon networks [37].

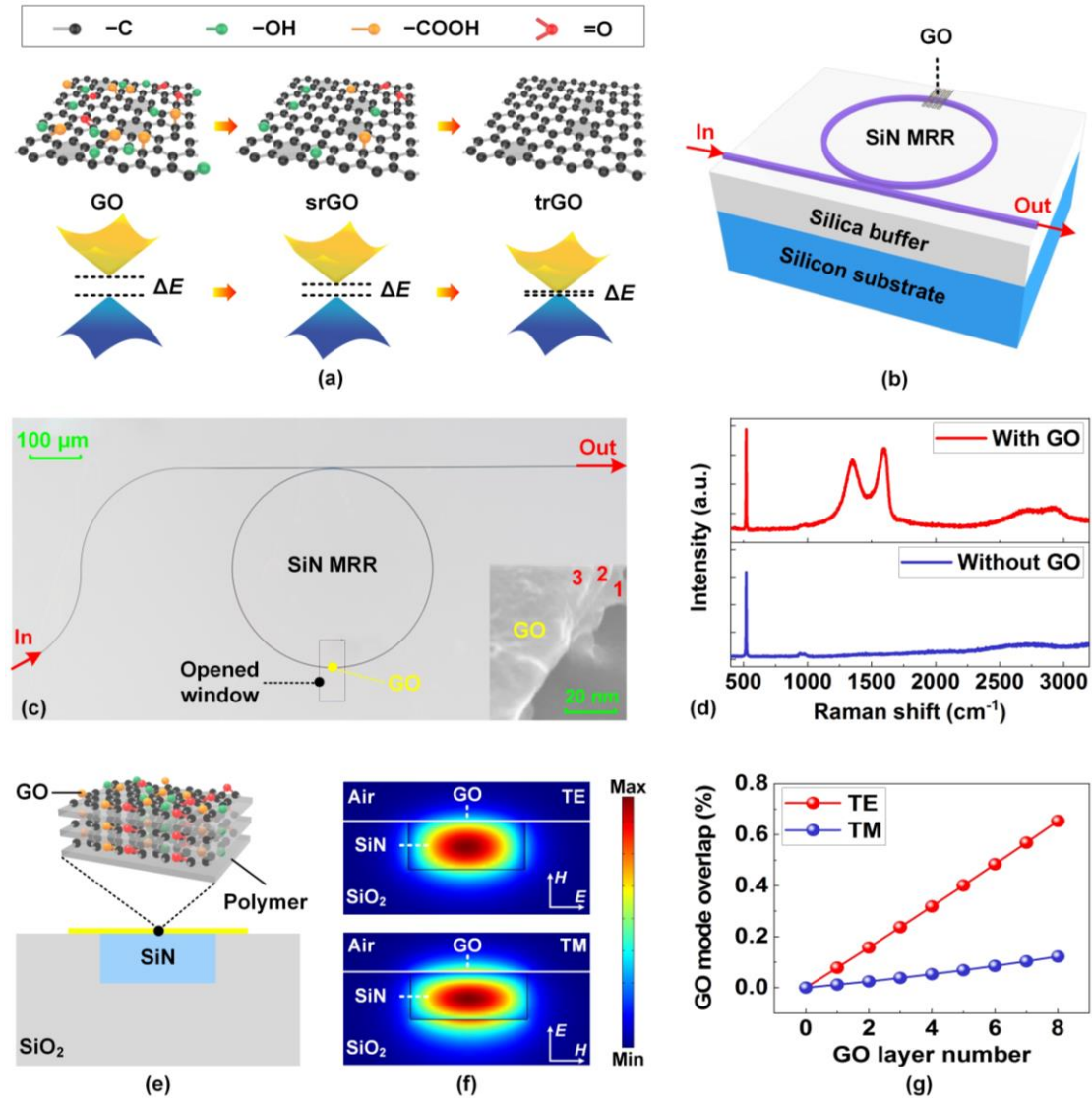

**Figure 1.** (**a**) Schematics of atomic structures and bandgaps of graphene oxide (GO), semi-reduced GO (srGO), and totally reduced GO (trGO). (**b**) Schematic of a GO-coated silicon nitride (SiN) microring resonator (MRR). (**c**) Microscopic image of a fabricated SiN MRR coated with 3 layers of GO. Inset shows a scanning electron microscopy (SEM) image of the layered GO film, where numbers (1–3) refer to the number of layers for that part of the image. (**d**) Measured Raman spectra of a SiN chip without GO and coated with 1 layer of GO. (**e**) Schematic illustration of cross section and (**f**) corresponding TE and TM mode profiles for the hybrid waveguide with 3 layers of GO. Inset in (**e**) illustrates the layered GO film fabricated by self-assembly. (**g**) Mode overlap with GO versus GO layer number for both TE and TM polarizations of the hybrid waveguides.

**Figure 1b** shows a schematic of an integrated SiN microring resonator (MRR) coated with a 2D GO film. **Figure 1c** shows a microscopic image of a fabricated device coated with a film including 3 layers of GO, with the inset showing a scanning electron microscopy (SEM) image of the layered GO film. The SiN MRR was fabricated via CMOS compatible and crack-free processes, as reported previously [29, 38]. A 2.3-μm-thick silica layer was deposited on the top of the fabricated SiN MRRs as an upper cladding. To facilitate the interaction between the GO film and the evanescent field of the SiN MRR, lithography and dry etching processes were employed to open a window in the silica cladding, extending down to the top surface of the SiN MRR. The coating of the 2D layered GO film was realized by using a solution-based method that allowed transfer-free and layer-by-layer film deposition [32, 39]. During the coating process, four steps were repeated to assemble a multilayered film consisting of alternating GO monolayers and polymer layers with oppositely charged surfaces. As can be seen from **Figure 1c,** the coated GO film exhibits high transmittance and good morphology. The film also shows a high uniformity without any noticeable wrinkling or stretching. The thickness for the self-assembled GO film increases almost linearly with the number of GO layers, with an average thickness of ~2 nm for each layer. Our GO coating method enables large-area film coating with precisely controlled film thickness, which was also used for on-chip integration of 2D GO films to realize other functional devices [17, 27, 29, 31, 34].

Figure 1d shows the measured Raman spectra for the SiN chip before and after coating 1 layer of GO, which were measured using a ~514-nm pump laser. The presence of the representative $D$ (1345 cm$^{-1}$) and $G$ (1590 cm$^{-1}$) peaks of GO in the Raman spectrum for the GO-coated chip confirms the successful integration of the GO film onto the chip.

In our fabricated SiN MRRs, the cross section for the waveguides forming the MRR (includes both the ring and the bus waveguide) was ~1.60 μm × ~0.72 μm, as illustrated in **Figure 1e**. The inset shows a schematic of the layered GO film fabricated by self-assembly. **Figure 1f** shows the TE and TM mode profiles for the hybrid waveguide with 3 layers of GO in **Figure 1e**, which were simulated using a commercial mode solving software. The corresponding effective refractive indices at 1550 nm were $n_{eff,\ TE}$ =

~1.784 and $n_{eff,\,TM}$ = ~1.713. In our simulation, the refractive indices of GO and SiN at 1550 nm were $n_{GO}$ = 1.972 and $n_{SiN}$ = 1.990, respectively. These values were obtained from our experiments introduced in the following sections. **Figure 1g** shows GO mode overlap versus GO layer number for both TE and TM polarizations, which were calculated based on mode simulations of the hybrid waveguide at 1550 nm. Due to the difference in volume between the bulk SiN waveguide and the ultrathin 2D GO films, most light power is confined within the SiN waveguide (> 88%), and the GO mode overlap is < 1%. For both TE and TM polarizations, the GO mode overlap increases with the GO layer number, mainly resulting from the increase in the GO film thickness. The GO mode overlap for TE polarization is higher than that for the TM polarization. Given that the GO film has much higher light absorption compared to the SiN waveguide, a stronger mode overlap with GO leads to a higher propagation loss for the GO-SiN hybrid waveguide.

## 3. Refractive indices and extinction coefficients

By using the fabricated MRRs coated with 2D GO films, we first characterize the refractive indices ($n$'s) and extinction coefficients ($k$'s) of the GO films with different thicknesses or after experiencing different degrees of reduction. For all the experiments performed in this section, we employed SiN MRRs with a radius of ~200 μm. The length of the coated GO films, which approximately equalled to the width of the opened window in the silica cladding, was ~50 μm. Light coupling to the fabricated devices was achieved by using lensed fibres that were butt-coupled to inverse-taper couplers located at both ends of the bus waveguides. The fibre-to-chip coupling loss was ~4 dB per facet. We also employed a polarization controller (PC) to adjust the polarization of the input light and characterized the $n$, $k$ values for both TE and TM polarizations.

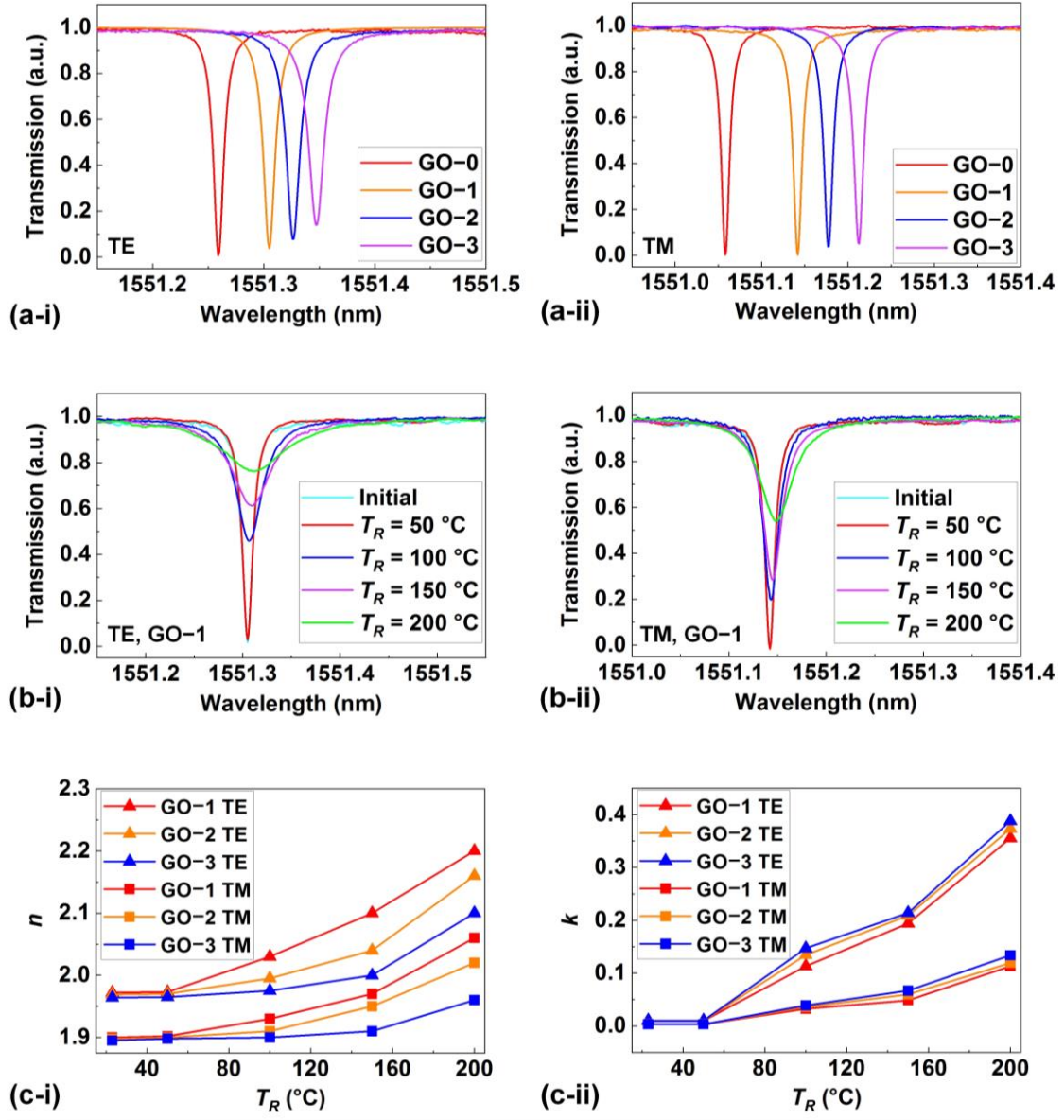

**Figure 2.** (**a**) Measured (**i**) TE- and (**ii**) TM-polarized transmission spectra of the hybrid MRRs with 1−3 layers of GO. The corresponding results for the uncoated SiN MRR (GO-0) are also shown for comparison. (**b**) Measured (**i**) TE- and (**ii**) TM-polarized transmission spectra of a hybrid MRR with 1 layer of GO. The GO-SiN chip underwent heating at various temperatures $T_R$ prior to the measurement of its transmission spectra. The corresponding results for the chip before heating (initial) are also shown for comparison. (**c**) Extracted (**i**) refractive index $n$ and (**ii**) extinction coefficient $k$ of the films including 1−3 layers of GO versus $T_R$ for both TE and TM polarizations.

**Figure 2a** shows the measured transmission spectra of a SiN MRR coated with 1−3 layers of GO for both TE and TM polarizations. The corresponding results for the uncoated SiN MRR (GO-0) are also shown for comparison. In order to minimize the wavelength shift induced by thermo-optic effects, the transmission spectra were measured by scanning the wavelength of an input continuous-wave (CW) light with a low power of ~0 dBm. Unless otherwise specified, the input power in our following discussion refers to the power coupled into the bus waveguide after excluding the fibre-to-chip coupling loss. Compared to the uncoated SiN MRR, the GO-coated MRRs exhibited resonance redshifts, which became more significant as the GO layer number increases. This indicates the variation in the waveguide effective refractive index caused by the GO films. On the other hand, the GO-coated MRRs exhibited decreased extinction ratios (*ER*s, defined as the ratio of the maximum to the minimum transmission) compared to the uncoated MRR, and the *ER* decreased for increasing GO layer numbers. This reflects that the presence of the GO films also resulted in changes to the waveguide propagation loss.

**Figure 2b** shows the measured transmission spectra of a SiN MRR coated with 1 layer of GO for both TE and TM polarizations. We measured the transmission spectra for the same MRR by scanning a CW light with a power of ~0 dBm (*i.e.*, the same as that in **Figure 2a**). Before measuring the spectra, the integrated chip was heated on a hot plate for 15 minutes at various temperatures $T_R$ ranging from ~50 to 200 °C. We did not further increase $T_R$ beyond 200 °C because the polymer layers within the GO films could not withstand temperatures in this range. The corresponding results for the unheated MRR (initial) are also shown for comparison. At $T_R$ = 50 °C, the spectrum of the heated MRR showed negligible difference compared to that of the unheated MRR. Whereas for $T_R \geq 100$ °C, the heated MRR exhibited a significantly decreased *ER*. Given that reduced GO has higher light absorption than unreduced GO [24], these results indicate that the 2D GO film did not undergo reduction at a relatively low $T_R$. However, as the temperature continued to rise above a certain threshold, thermal reduction of GO occurred, and the degree of reduction increased with $T_R$. On the other hand, we observed very slight redshifts in the resonance wavelength for $T_R \geq 100$ °C. This reflects that the reduction of GO also resulted in slight changes in the waveguide effective refractive index.

**Figure 2c** shows *n*, *k* values of the layered GO films, which were obtained by using the scattering matrix method [40] to fit the measured transmission spectra in **Figures 2a** and **b**. The refractive index of SiN was ~1.9902 at 1550 nm, which was obtained by fitting the measured spectra for the uncoated SiN MRR. To improve the accuracy, the slight decrease in the film thicknesses (*i.e.*, ~0.5 nm for 1 layer of GO at $T_R$ = 200 °C, measured by atomic force microscopy) was considered when we calculated the *n*, *k* values for reduced GO. In **Figure 2c**, *n* slightly decreases as the GO layer number increases, but *k* shows an opposite trend. The former could be attributed to the presence of more air voids in a thicker GO film, which arose from imperfections during the GO film's fabrication processes. Whereas the latter was resulting from an increase in scattering loss due to film unevenness and accumulation of imperfect contact between adjacent layers in a thicker film.

In **Figure 2c**, both *n* and *k* initially exhibited no significant changes for $T_R \leq 50$ °C, but then show a more noticeable increase with $T_R$ for $T_R \geq 100$ °C. These results reflect the changes in GO's *n*, *k* induced by thermal reduction. Since the CW light power used to scan the spectra was not sufficient to induce any significant thermo-optic effects, the changes in *n*, *k* were mainly induced by changes resulting from the heating treatment before the measurement. The *n*, *k* values for unreduced GO were ~1.9720 and ~0.0097, respectively. In contrast, the corresponding values for the reduced GO at $T_R$ = 200 °C were ~2.2000 and ~0.3557. The change in *n*, *i.e.*, $\Delta n$ = ~0.2280, is over 1 order of magnitude larger than conventional bulk refractive materials [41]. In addition, *k* also changes from ~0.0097 to ~0.3557 – an increase of over 36-fold. The broad variation ranges for both *n*, *k* are critical for engineering and manipulation of the phase and amplitude response of many GO devices, such as optical lenses and holographic displays [31, 42].

The results in **Figure 2c** also reveal the differences in the *n*, *k* values between TE and TM polarizations. For all different GO layer numbers and $T_R$'s, the GO film exhibits higher values of *n*, *k* for TE polarization than TM polarization. This is mainly caused by the intrinsic material anisotropy of the 2D GO films with atomically thin thicknesses, which leads to distinctive response for light in different polarization states. In the GO-SiN MRR, the TE polarization supports in-plane light-GO interaction, which is much stronger as compared to the out-of-plane interaction supported by TM polarization. For 1 layer of GO, the $n$, $k$ values for TE polarization were ~1.9720 and ~0.0097, respectively. In contrast, the corresponding values for TM polarization were ~1.9000 and ~0.0032. The large difference in the $n$, $k$ values between TE and TM polarizations reflect the significant material anisotropy of the 2D GO films, which is useful for realizing liquid crystal display devices and optical polarizers [28, 43].

## 4. Thermo-optic coefficients

In this section, we use the fabricated GO-SiN MRRs to characterize the thermo-optic coefficients of the GO films with different thicknesses or after experiencing different degrees of reduction. The thermo-optic coefficient is a fundamental parameter that indicates the change in the refractive index of a material with variations in environmental temperature. For the experiments performed in this section, we employed SiN MRRs with a radius of ~56 µm that was smaller than those used in **Section 3**. In addition, the silica upper cladding of the integrated chip was removed by using highly selective chemical-mechanical polishing (CMP) and the entire MRR was coated with GO films. We chose these MRRs in order to magnify the difference in measured wavelength shifts caused by GO films. We measured the transmission spectra of the GO-SiN MRRs by scanning the wavelength of a CW input with a power of ~0 dBm. To adjust the temperature of the GO-SiN MRRs, the fabricated chip was placed on a temperature controller with a minimum resolution of 0.1 °C.

**Figure 3a** shows the measured resonance wavelength $\lambda_{res}$ versus chip temperature $T$ for the SiN MRRs coated with 1−3 layers of GO. We show the results for both TE and TM polarizations, and the corresponding results for the uncoated SiN MRR (GO-0) are also shown for comparison. To prevent changes in GO film properties caused by thermal reduction, the chip temperature was not raised beyond 50 °C. As can be seen, all the MRRs exhibited a redshift in their resonance wavelengths, albeit at slightly different rates. For instance, the TE-polarized $\lambda_{res}$ of the uncoated MRR redshifted at a rate of ~22.9 pm/°C, in contrast to ~23.3 pm/°C for the TE-polarized $\lambda_{res}$ of the hybrid MRR with 1 layer of GO. **Figure 3b** shows the measured $\lambda_{res}$ versus $T$ for a hybrid MRR with 1 layer of GO. Similar to that in **Figure 2b**, before conducting the measurement, the GO-SiN chip was heated at various $T_R$ on a hot plate for 15 minutes. As a result, the measured redshifts in **Figure 3b** can be used to calculate the thermo-optic coefficients of reduced GO at various reduction degrees.

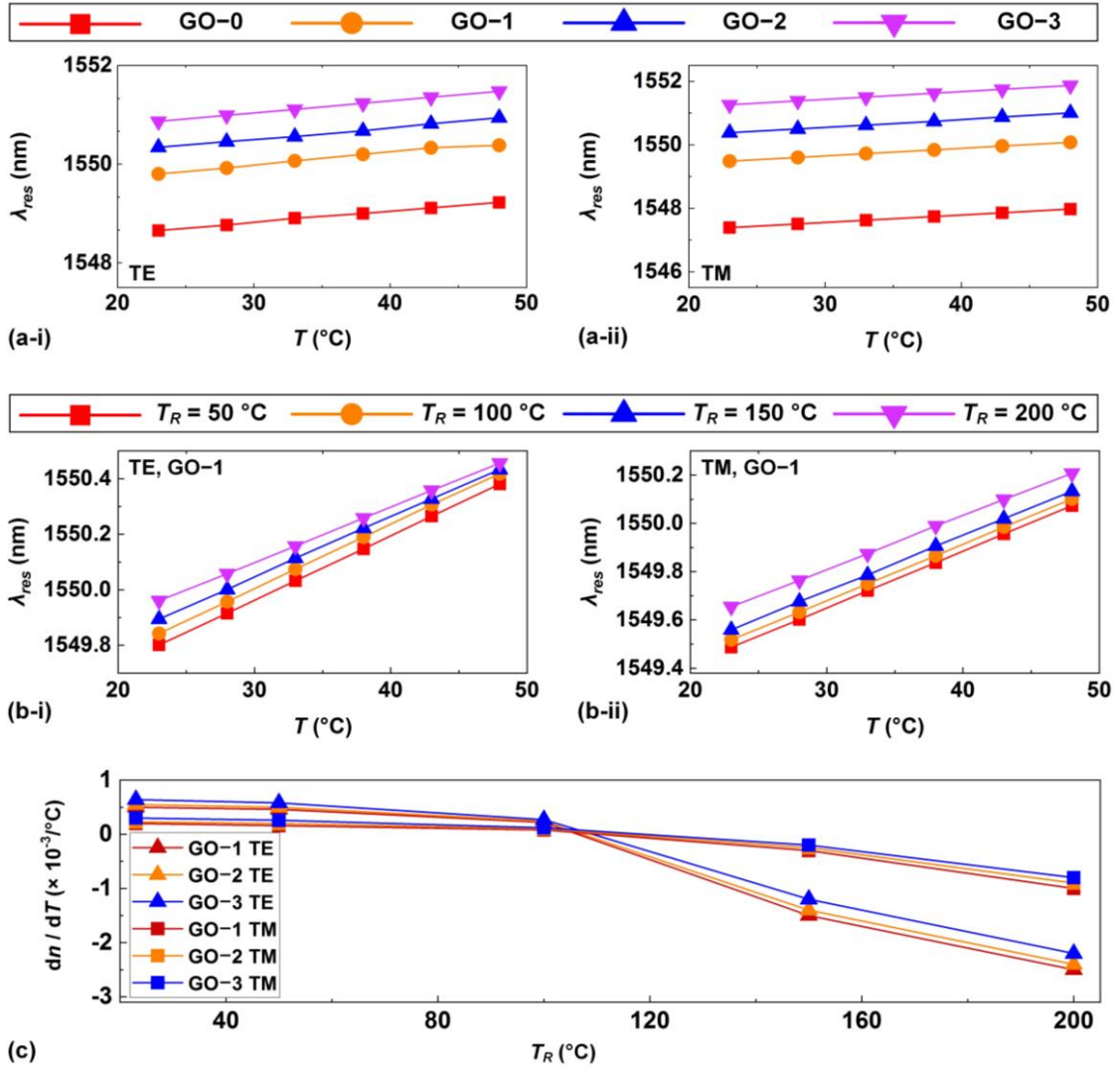

**Figure 3.** (**a**) Resonance wavelength $\lambda_{res}$ versus chip temperature $T$ (changed via a temperature controller) for the hybrid MRRs with 1−3 layers of GO. (**i**) and (**ii**) show the results for TE and TM polarizations, respectively. The corresponding results for the uncoated SiN MRR (GO-0) are also shown for comparison. (**b**) $\lambda_{res}$ versus $T$ for the hybrid MRR with 1 layer of GO. The GO-SiN chip underwent heating at temperatures $T_R$ ranging from ∼50 to 200 °C prior to the measurement. (**i**) and (**ii**) show the results for TE and TM polarizations, respectively. (**c**) Extracted thermo-optic coefficient d$n$ / d$T$ of the films including 1−3 layers of GO versus $T_R$ for both TE and TM polarizations.

**Figure 3c** shows thermo-optic coefficient d$n$ / d$T$ of the films including 1−3 layers of GO versus $T_R$ for both TE and TM polarizations. The values of d$n$ / d$T$ were calculated based on the results in **Figures 3(a)** and **(b)**, using the relationship between the resonance wavelengths and the waveguide effective refractive index given by:

$$n_{eff} \times 2\pi / \lambda_m \times L = m \times 2\pi \tag{1}$$

where $n_{eff}$ is the effective refractive index of the hybrid waveguide, $L$ is the circumference of the SiN MRR, and $\lambda_m$ is the resonance wavelength of the $m$th resonance.

In **Figure 3c**, as $T_R$ increases, the value of $dn / dT$ changes from positive to negative for all the three GO layer numbers. For 1 layer of unreduced GO in TE polarization, the value of $dn / dT$ is ~$0.5 \times 10^{-3}$/°C, in contrast to ~$-2.5 \times 10^{-3}$/°C for reduced GO at $T_R$ = 200 °C with the same GO layer number and polarization. These results reflect interesting changes in GO's thermo-optic coefficient induced by thermal reduction. The thermo-optic coefficient of unreduced GO is higher than silicon (~$1.8 \times 10^{-4}$/°C [44]). Given that most materials have positive values of thermo-optic coefficients, the negative thermo-optic coefficients exhibited by reduced GO can be utilized to mitigate thermal drift induced by temperature variation and develop athermal devices [40, 45]. The large variation range for the thermo-optic coefficient also increases the range of functionalities and devices that can be developed.

We also note that thicker GO films showed higher values of $dn / dT$ in **Figure 3c**, which can be attributed to diminished thermal dissipation in them, especially considering the presence of polymer layers between the GO layers. Similar to that in **Figure 2c**, the GO film exhibits significant difference in the values of $dn / dT$ for TE and TM polarizations. For unreduced GO with positive values of $dn / dT$, the $dn / dT$ for TE polarization is higher than that for TM polarization. Whereas for reduced GO at $T_R$ = 200 °C with negative values of $dn / dT$, the $dn / dT$ for TE polarization is lower than that for TM polarization. These phenomena can also be attributed to the stronger in-plane light-GO interaction supported by TE polarization compared to the out-of-plane interaction supported by TM polarization. We also calculated the $dn / dT$ values of SiN based on the measured results for an uncoated SiN MRR, which were ~2.5 × $10^{-5}/°C$ and ~$2.4 \times 10^{-5}/°C$ for TE and TM polarizations, respectively. These values show agreement with those reported in Ref. [46], and the close resemblance between the coefficients for TE and TM polarizations reflects that SiN does not exhibit significant anisotropy in terms of its thermo-optic coefficient.

## 5. Thermal conductivities

Thermal conductivity is a parameter that defines material's ability to conduct heat, which is of fundamental importance in modelling heat transfer for thermal management [18]. In this section, we characterize the thermal conductivities of the layered GO films with different thicknesses and reduction degrees by using the fabricated GO-SiN MRRs.

For the experiments performed in this section, we employed SiN MRRs with a radius of ~200 μm, and the length of the coated GO films was ~50 μm (*i.e.*, the same as those used in **Section 3**). We used two CW lights to measure the transmission spectra of the GO-coated MRRs. The first one served as a pump injecting into one of the MRR's resonances. The wavelength of this input CW light was slightly tuned from blue to red around the resonance until it reached a steady thermal equilibrium state with a stable output power. After this, the second CW light, with a constant power of ~0 dBm, was employed as a low-power probe to scan the MRR's transmission spectrum. Compared to directly employing a high-power CW light to scan the spectra, our approach can minimize the asymmetry in the measured resonance lineshape resulting from optical bistability (which will be discussed in **Section 7**), thus allowing more precise measurement of the resonance wavelength shifts. We measured the resonance wavelength shifts of the fabricated GO-SiN MRRs for various input pump powers, and extracted the thermal conductivity by fitting the wavelength shift with theoretical simulations.

**Figure 4a** shows the measured resonance wavelength shift $\Delta\lambda$ versus the power of the CW pump $P_p$ for the uncoated SiN MRR and the hybrid MRR with 1 layer of GO. We chose the range of $P_p \leq$ ~40 mW because, within this range, there were no observable changes in the $ER$ of the hybrid MRR compared to that measured at $P_p = 0$. This implies that GO film did not experience thermal reduction induced by the CW pump power. For both the uncoated and the hybrid MRRs, $\Delta\lambda$ increases gradually for smaller values of $P_p$, and then increases more significantly as $P_p$ increases. As $P_p$ increases, the difference in $\Delta\lambda$ between the uncoated and hybrid MRRs also becomes more significant.

In our modelling, the steady-state temperature distributions in the waveguide cross section were simulated using commercial finite-element multi-physics software (COMSOL Multiphysics) to solve the heat equation described by the *Fourier*'s law as [18]:

$$-\nabla \cdot (K \nabla T) = q \tag{2}$$

where $q$ is the heat flux density, $K$ is the material's thermal conductivity, $\nabla T$ is the temperature gradient, and $\nabla$ acting on the vector function $K \nabla T$ is the divergence operator. The heat flux intensity $q$ in the MRR was calculated by [47]:

$$q = \frac{P \times BUF \times R}{w \times h \times l} \tag{3}$$

where $P$ is the input CW power that equals to $P_p$ (neglecting the small difference induced by the low-power probe), $R$ is the conversion efficiency that determines the amount of light power converted into heat, $w$, $h$, and $l$ are the waveguide width, height, and length, respectively, and $BUF$ is the MRR's intensity build-up factor that can be expressed as [48]:

$$BUF = \frac{(1 - t^2)a^2}{1 - 2ta + t^2a^2} \tag{4}$$

where $a$ is the round-trip amplitude transmission, and $t$ is the field transmission coefficient of the directional coupler. The values of $a$ and $t$ were obtained by fitting the measured spectrum of the MRR.

Based on the measured resonance wavelength shift for the uncoated SiN MRR in **Figure 4a**, we first calculated the variation in the waveguide effective refractive index, and extracted the change in the refractive index of SiN $\Delta n_{SiN}$ based on optical mode simulations (*e.g.*, $\Delta n_{SiN} = \sim 8.54 \times 10^{-5}$ at $P_p = 40$ mW). Next, by dividing $\Delta n_{SiN}$ by the thermo-optic coefficient of SiN (*i.e.*, ~2.50 × 10$^{-5}$/°C, obtained from the experiments in **Section 4**), we derived the change in temperature $\Delta T$ at the waveguide core (*e.g.*, $\Delta T$ = ~3.4 °C at $P_p$ = 40 mW). Finally, by fitting the $\Delta T$ value at the waveguide core with simulated temperature distribution, we obtained the heat flux intensity $q$ and calculated $R$ for the uncoated SiN MRR based on **Eq. (3)** (*i.e.*, $R$ = ~0.66%).

On the other hand, we divided the hybrid MRR into uncoated and GO-coated segments. For the uncoated segment, since the presence of GO increased the round-trip loss of the MRR and hence reduced the *BUF*, the heat flux density and temperature increase in the uncoated segment were lower compared to those for the uncoated SiN MRR. Assuming that $R$ for the uncoated segment remains consistent with that obtained from the uncoated MRR, and neglecting the slight temperature variation induced by heat transfer between the uncoated and GO-coated segments, we obtained the $\Delta T$ for the uncoated segment in the hybrid MRR via thermal simulations (using $q$ calculated based on **Eq. (3)**). Using this $\Delta T$, we further calculated the additional phase shift along the uncoated segment based on **Eq. (1)**. This, together with the measured resonance wavelength shift for the hybrid SiN MRR in **Figure 4a**, allows us to calculate the additional phase shift along the GO-coated segment and the corresponding $\Delta T$ in this segment.

Based on the methods mentioned above, we calculated $\Delta T$ values for the uncoated and GO-coated segments of a hybrid MRR with 1 layer of GO, and plotted them as functions of $P_p$ in **Figure 4b**. It is evident that, owing to the presence of GO, the two segments experienced different temperature changes at the SiN waveguide core. Given that the hybrid MRR with 1 layer of GO has a *BUF* value of ~23.0 that is lower than the *BUF* of ~28.2 for the uncoated MRR, the uncoated segment within the hybrid MRR has a reduced heat flux intensity than that in the uncoated MRR. This leads to a lower $\Delta T$ value for the uncoated segment within the hybrid MRR relative to the uncoated MRR (*e.g.*, ~2.8 °C versus ~3.4 °C at $P_p$ = 40 mW). Even though the lower $\Delta T$ caused a smaller phase shift along the uncoated segment, in **Figure 4a** the hybrid MRR exhibited a more significant redshift compared to the uncoated MRR. This indicates that the GO-coated segment experienced a more significant phase shift resulting from the heat generated by the GO films. This is also reflected by the higher values of $\Delta T$ for the GO-coated segment in **Figure 4b**.

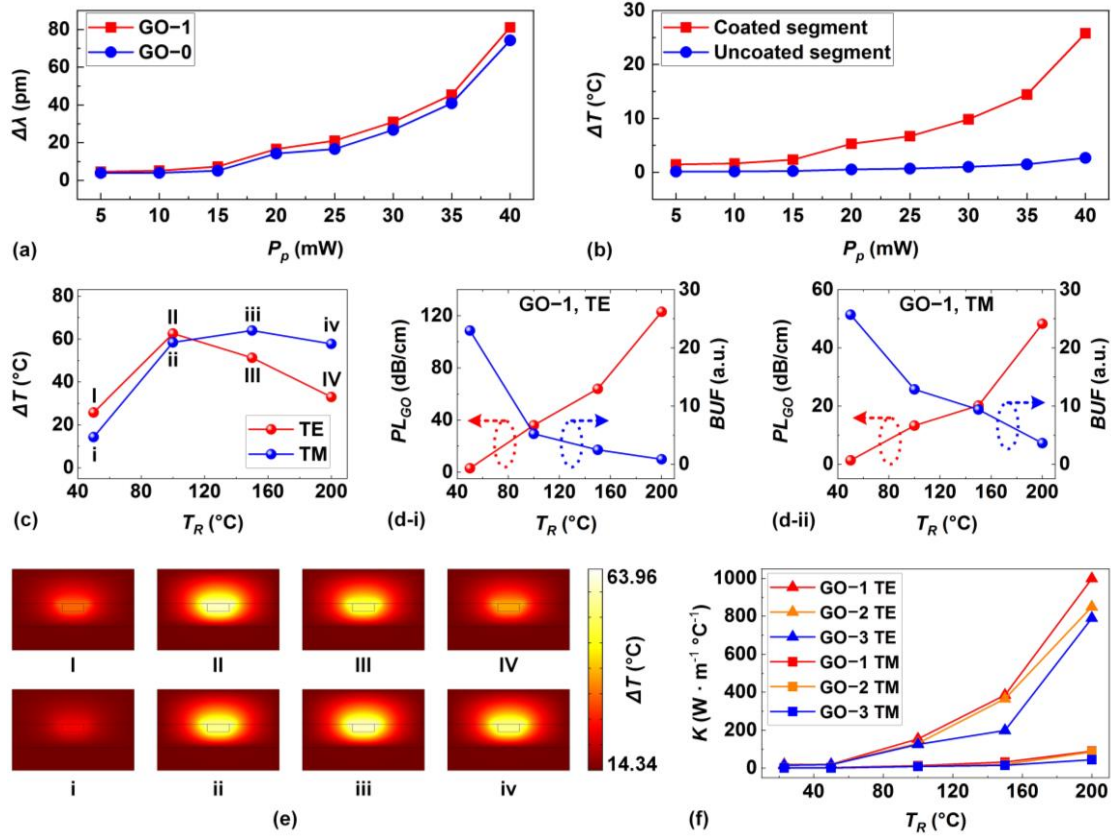

**Figure 4.** (**a**) Measured resonance wavelength shift $\Delta\lambda$ versus input continuous-wave (CW) pump power $P_p$ for the uncoated SiN MRR (GO-0) and the hybrid MRR with 1 layer of GO (GO-1). (**b**) Calculated changes at the waveguide core $\Delta T$ for the uncoated and GO-coated segments in the hybrid MRR with 1 layer of GO versus $P_p$. (**c**) Calculated $\Delta T$ for the GO-coated segment in the hybrid MRR with 1 layer of GO after the chip was heated at various temperatures $T_R$. (**d**) Calculated GO-induced propagation loss $PL_{GO}$ and intensity build up factor $BUF$ of the hybrid MRR with 1 layer of GO versus $T_R$. (**i**) and (**ii**) show the results for TE and TM polarizations, respectively. (**e**) Simulated temperature distributions in the waveguide cross section for the hybrid MRR with 1 layer of GO. (I) – (IV) and (i) – (iv) correspond to the different points in (**c**). (**f**) Extracted thermal conductivity $K$ of the films including 1−3 layers of GO versus $T_R$ for both TE and TM polarizations. In (**c**) – (**e**), $P_p$ = ~40 mW, at which there was no observable changes in the extinction ratios of the hybrid MRRs as compared to those measured at $P_p$ = 0. In the thermal simulation, the initial temperature was assumed to be at room temperature of 23 °C.

After measuring the resonance wavelength shifts of a hybrid MRR with 1 layer of GO following the heating of the chip at various $T_R$, we employed the same method as in **Figure 4b** to calculate $\Delta T$ for the GO-coated segment in these MRRs. **Figure 4c** shows the calculated $\Delta T$ versus $T_R$ for both TE and TM polarizations. For comparison, the input pump power was kept the same as $P_p = \sim 40$ mW, at which there were no observable changes in the $ERs$ of these MRRs as compared to those measured at $P_p = 0$. For both TE and TM polarizations, $\Delta T$ initially rises with $T_R$ and then starts to decline after reaching peak values at $T_R = 100$ °C. To delve deeper into the underlying reasons for this interesting phenomenon, we calculated the GO-induced propagation loss $PL_{GO}$ and the $BUF$ of these MRRs, and depicted them in **Figure 4d**. As can be seen, $PL_{GO}$ increases with $T_R$, while $BUF$ exhibits an opposite trend. This is not surprising given the fact that the reduced GO at $T_R \geq 100$ °C has a higher light absorption than unreduced GO. Since the light absorbed by the GO film is converted to heat, the increased light absorption for reduced GO also leads to a higher value of $R$ in **Eq. (3)**. Hence, the observed phenomenon in **Figure 4c** can be attributed to the trade-off between a decreased $BUF$ and an increased $R$ in **Eq. (3)** as $T_R$ increases.

Based on the calculated $\Delta T$ values for the GO-coated segment, we further extracted the thermal conductivities of the GO films by fitting the $\Delta T$ values with those obtained from thermal simulations based on the finite-element method (FEM). In our simulations, the GO film in the GO-coated segment was regarded as an additional heat source that absorbed light power and generated heat, which in turn caused a temperature change, $\Delta T$, in the underlying SiN waveguide. The heat flux densities in the GO films and the SiN waveguide were calculated individually based on **Eq. (3)**. The values of $R$ for the GO films were calculated based on the linear loss of the GO films obtained in **Section 3**, assuming that all the light absorbed by GO – equivalent to the additional loss induced by GO – was converted to heat. The values of $K$ for each of the material regions were also specified (*e. g.*, $K_{SiN} = \sim 29.00$ W/(m·°C), $K_{SiO2} = \sim 1.40$ W/(m·°C), $K_{polymer} = \sim 0.25$ W/(m·°C), and $K_{air} = \sim 0.03$ W/(m·°C) [47, 49]), with the exception of the GO film that needed fitting. The initial temperature $T_0$ was set to 23 °C, which was the ambient temperature during our experiments.

**Figure 4e** shows the simulated temperature distributions in the waveguide cross section for the hybrid MRR with 1 layer of GO, where (I) – (IV) and (i) – (iv) correspond to the different points in **Figure 4c**. **Figure 4f** shows the thermal conductivity $K$ of the films including 1−3 layers of GO versus $T_R$ for both TE and TM polarizations. With the exception of the results for 2 and 3 layers of unreduced GO in TE polarization, where the fitting was based on measurements at $P_p = \sim 30$ mW due to changes in the $ER$ observed at $P_p \geq \sim 35$ mW, all other results were obtained by fitting the measurements at $P_p = \sim 40$ mW.

In **Figure 4f**, a thicker GO film exhibits a lower value of $K$, further confirming its lower thermal dissipation property. For 1 layer of unreduced GO, the value of $K$ is ~20.6 W/(m·°C) for TE polarization. This value is higher than that reported in Ref. [50] (*i.e.*, ~2.0 W/(m·°C)) for GO films with greater thicknesses (*e.g.*, ~143 nm). The decrease in the thermal conductivity with increasing film thickness can be attributed to more significant scattering of phonons from the film surface caused by film unevenness, lattice impurities, and structural defects within thicker films.

Compared to unreduced GO, higher values of $K$ are obtained for reduced GO at $T_R \geq 100$ °C, with $K$ increasing as the GO reduction degree increases. For 1 layer of reduced GO at $T_R = 200$ °C, the value of $K$ is ~998 W/(m·°C) for TE polarization, which is more than 48 times that of comparable unreduced GO. These results reflect that the reduced GO films possess superior heat conduction property, which can be attributed to the diminished presence of the OFGs that induce phonon-scattering from the film surface. We also note that the value of ~998 W/(m·°C) is about five times lower than the reported value for graphene [51], which is possibly due to the fact that the GO film at $T_R = 200$ °C was still not fully reduced and the existence of polymer layers with a much lower thermal conductivity in our GO films.

We also note that the TE polarization exhibited higher values of $K$ than the TM polarization for all the three GO layer numbers in **Figure 4f**. This indicates that the GO films also exhibit anisotropic thermal conductivity – similar to the anisotropy observed in their refractive index, extinction coefficient, and thermo-optic coefficient. The ratios between the in-plane and out-of-plane thermal conductivities are ~6.6 for 1 layer of unreduced GO and ~11.1 for 1 layer of reduced GO at $T_R = 200$ °C. We note that in Ref. [52] a considerably higher ratio of ~675 was achieved for thick (~40 μm) reduced GO films annealed at 1000 °C. In these films, the formation of "air pockets" with limited thermal transport capability between the GO layers strongly hindered out-of-plane thermal transport, but not significantly affecting the in-plane thermal conduction. In contrast to this ratio achieved through engineering the fabrication processes of thick GO films, our calculated ratios more accurately reflect inherent anisotropy in the thermal conductivities of 2D GO films.

## 6. GO reduction induced by localized photo-thermal effects

For the experiments in **Section 5**, the input CW pump power $P_p$ was maintained below a specific threshold to ensure that there were no obvious changes in the *ERs* of the hybrid MRRs compared to that measured at $P_p = 0$. As a result, the GO films remained unaffected by thermal reduction induced by the CW pump power. In this section, we further increase $P_p$ to be above the threshold and characterize the changes in the *ERs* of the hybrid MRRs induced by thermal reduction of the GO films. Compared to thermal reduction of GO caused by heating the entire chip on a hotplate (at various $T_R$) in previous experiments, using the CW pump power to trigger thermal reduction of GO leads to localized effects, which can cause dynamic changes of the GO film properties. For the experiments in this section, we employed the same experimental method as well as the same MRRs as those used in **Section 5**.

**Figure 5a** shows the measured *ER* versus $P_p$ for the hybrid MRR with 1 layer of GO. In (i) and (ii), we show the results for TE and TM polarizations, respectively. In **Figure 5a-i**, the *ER* remained constant at ~14.0 dB for $P_p \leq 40$ mW. However, an obvious decrease in the *ER* was observed for $P_p > 40$ mW, indicating increased loss in the hybrid MRR induced by thermal reduction of the GO film. In the hybrid MRR, localized heating occurs due to optical absorption, which leads to a range of photo-thermal effects within the GO film such as self-heating, thermal dissipation, and photo-thermal reduction. Upon reaching the power threshold, the reduction of GO can be initiated when the thermal heating initiates a deoxygenation reaction in GO. The reduced GO with higher light absorption increased the loss in the MRR, consequently lowering its *ER*. We also observed an interesting phenomenon – when $P_p$ exceeded lower power threshold of $P_{thres1}$ = ~40 mW but remained below higher power threshold of $P_{thres2}$ = ~72 mW (which is highlighted by the blue shaded area), the *ER* of the hybrid MRR returned to its initial value of ~14.0 dB upon switching off the CW pump. This reversibility reflects the instability of the reduced GO induced by photo-thermal effects, which can easily revert to the original unreduced state once it has cooled down in an oxygen-containing ambient. The reversible GO reduction within the MRR enables dynamic tuning of the MRR's *ER*, which can be useful for potential applications such as optical switches and power limiting [53, 54].

As $P_p$ continued to rise above ~72 mW, there were permanent changes in the *ER* after switching off the CW pump. This occurred because the temperature rise resulting from localized heating caused by the CW pump reached a level where it permanently fractured the chemical bonds between the OFGs and the carbon network. This led to a lasting alteration in the atomic structure of GO and hence its properties. For $P_p$ > 110 mW, the *ER* nearly approaches 0, indicating a substantial increase in loss induced by the highly reduced GO. It should be noted that for $P_p$ > ~72 mW, there was an increase in the *ER* after tuning off the CW pump, although the *ER* could not return to the initial value of ~14.0 dB. At $P_p$ = ~160 mW, there was still a slight increase in the *ER* (~0.2 dB) when the CW pump was off, reflecting the fact that the highly reduced GO at this power level was not yet fully reduced.

In **Figure 5a-i**, the dashed horizontal lines indicate the measured *ER*s for the hybrid MRR with 1 layer of GO after the chip was heated on a hotplate at various $T_R$ (*i.e.*, the same as those in **Figure 2b**, which were measured by scanning a CW light with a power ~1 mW). These lines were plotted to compare the performance between localized heating using a CW pump and uniform heating of entire chip via a hotplate. As can be seen, despite using different methods to reduce the GO film, similar *ER*s can be achieved for the hybrid MRRs with GO films reduced to different degrees.

For TM polarization in **Figure 5a-ii**, reversible GO reduction was observed for $P_p$ within the range of ~82 – 123 mW. Compared to TE polarization, the lower power threshold $P_{thres1}$ is higher and the power range for reversible reduction is larger for TM polarization. This reflects the relatively weak photo-thermal effects for TM polarization due to the significant anisotropy of the GO film, aligning with the experimental results in previous sections.

**Figures 5b** and **5c** show the corresponding results for the hybrid MRRs with 2 and 3 layers of GO, respectively. For the hybrid MRR with 2 layers of GO, reversible GO reduction was observed for $P_p$ within the range of ~36 – 58 mW, and for the hybrid MRR with 3 layers of GO, it was observed for $P_p$ within the range ~32 – 47 mW. Compared to the results in **Figure 5a**, the hybrid MRR with a thicker GO film shows a decreased $P_{thres1}$ and a smaller power range, reflecting the more significant photo-thermal effects in them.

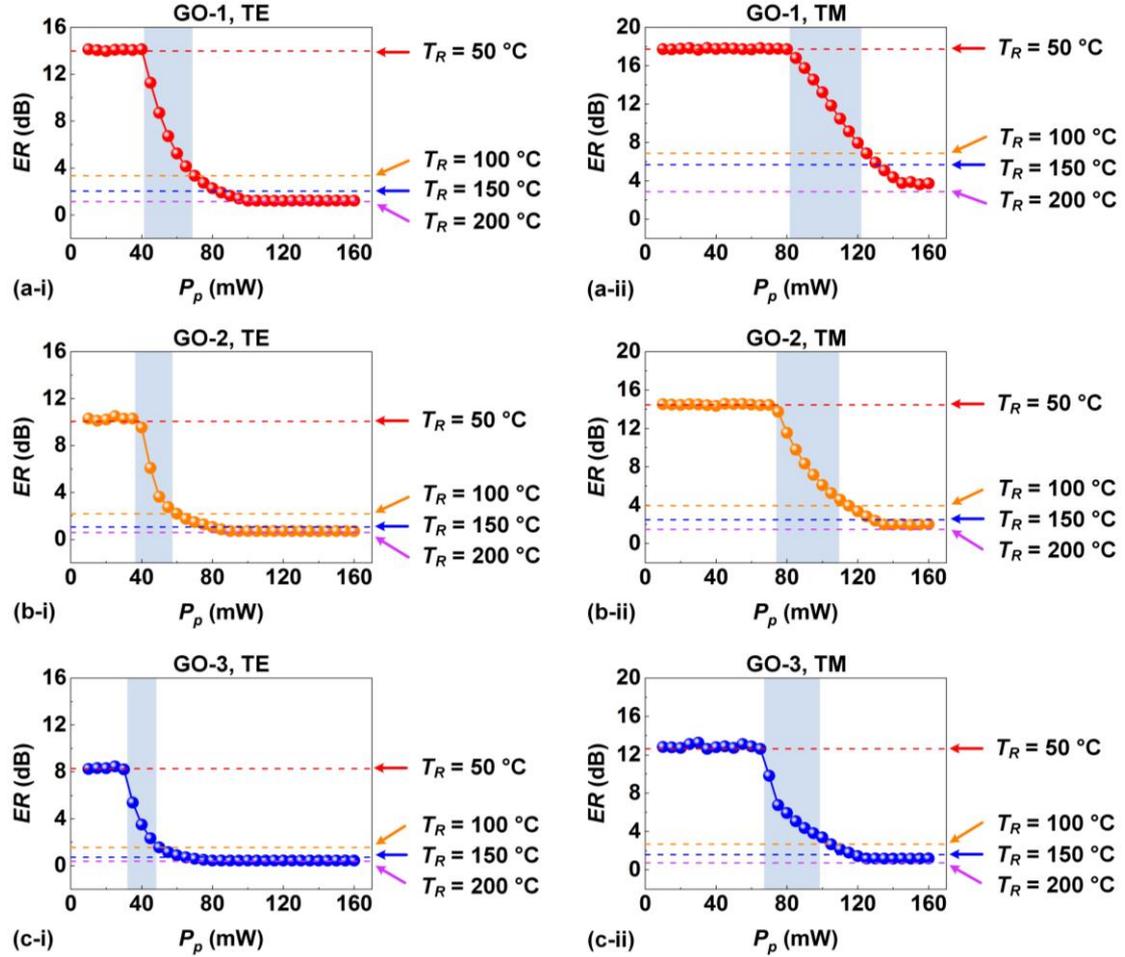

**Figure 5.** Measured extinction ratio $ER$ versus input CW pump power $P_p$ for the hybrid MRRs with (**a**) 1, (**b**) 2, and (**c**) 3 layers of GO. In (**a**) – (**c**), (**i**) and (**ii**) show the results for TE and TM polarizations, respectively. The blue shaded areas indicate the power ranges associated with reversible GO reduction. The dashed horizontal lines indicate the $ER$s for the corresponding MRRs after the chip was heated at various temperatures $T_R$ and measured by scanning a CW light with a power of ~1 mW.

In **Figure 5**, before injecting the CW pump, the hybrid MRRs were coated with unreduced GO films. In **Figure 6**, we further characterize the performance for the hybrid MRRs with reduced GO films before injecting the CW pump. In **Figures 6a** and **6b**, we show the measured $ER$ versus $P_p$ for the hybrid MRR with 1 layer of reduced GO for TE and TM polarizations, respectively. Prior to injecting the CW pump and measuring the transmission spectra, the chip was heated on a hotplate to obtain reduced GO at different reduction degrees. In each figure, (i) – (iii) show the results measured after the chip was heated at $T_R$ = ~100 °C, ~150 °C, and ~200 °C, respectively.

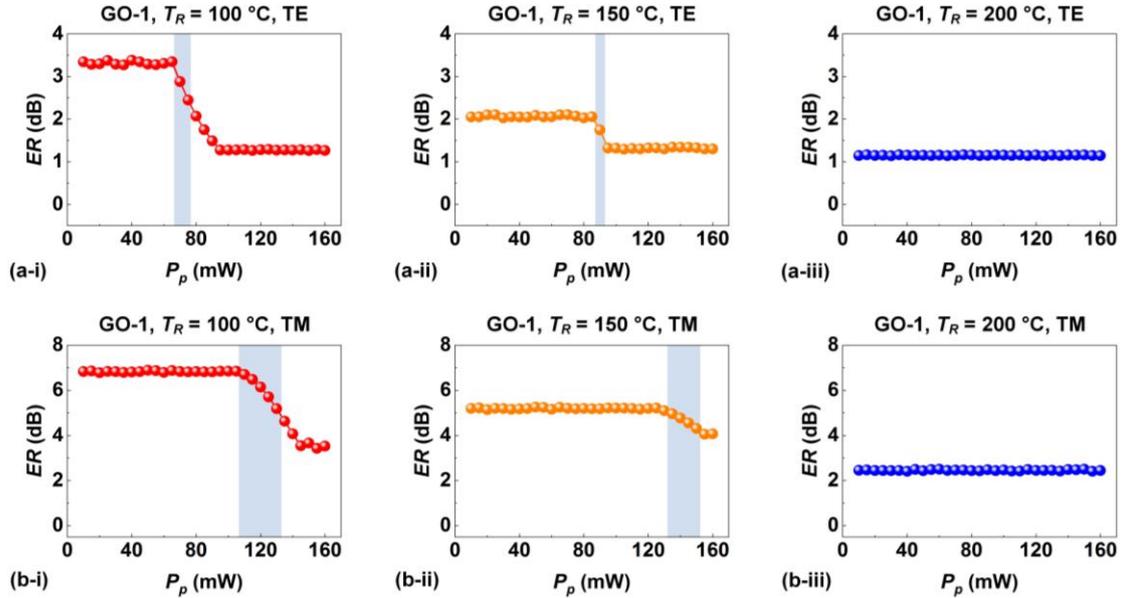

**Figure 6.** Measured extinction ratio $ER$ versus input CW pump power $P_p$ for the hybrid MRR with 1 layer of reduced GO for (**a**) TE and (**b**) TM polarizations. In (**a**) – (**b**), (**i**) – (**iii**) show the results measured after the chip was heated at temperatures of $T_R$ = ~100 °C, $T_R$ = ~150 °C, and $T_R$ = ~200 °C, respectively. The blue shaded areas indicate the power ranges associated with reversible GO reduction.

For $T_R$ = ~100 °C in **Figure 6a-i**, reversible GO reduction was observed for $P_p$ within the range of ~65 – 78 mW. Compared to the results for unreduced GO in **Figure 5a-i**, the $P_{thres1}$ is higher, and the power range is smaller. This indicates that unreduced GO was more easily reduced by the CW pump. In **Figure 6a-ii**, for reduced GO at $T_R$ = 150 °C, reversible GO reduction was observed with an even higher $P_{thres1}$ and in an even smaller range of ~87 – 93 mW. For the reduced GO at $T_R$ = 200 °C in **Figure 6a-** **iii**, no significant changes in the *ER* were observed, indicating that there was no obvious reversible GO reduction for highly reduced GO. These results further confirm that the reversible reduction behaviour becomes less obvious as the degree of reduction increases. Similar to **Figure 5**, the results for TM polarization in **Figure 6b** show higher $P_{thres1}$ values and larger ranges for reversible GO reduction compared to the corresponding results for TE polarization in **Figure 6a**. At $T_R$ = 200 °C, there were also no obvious changes in the *ER* for TM polarization.

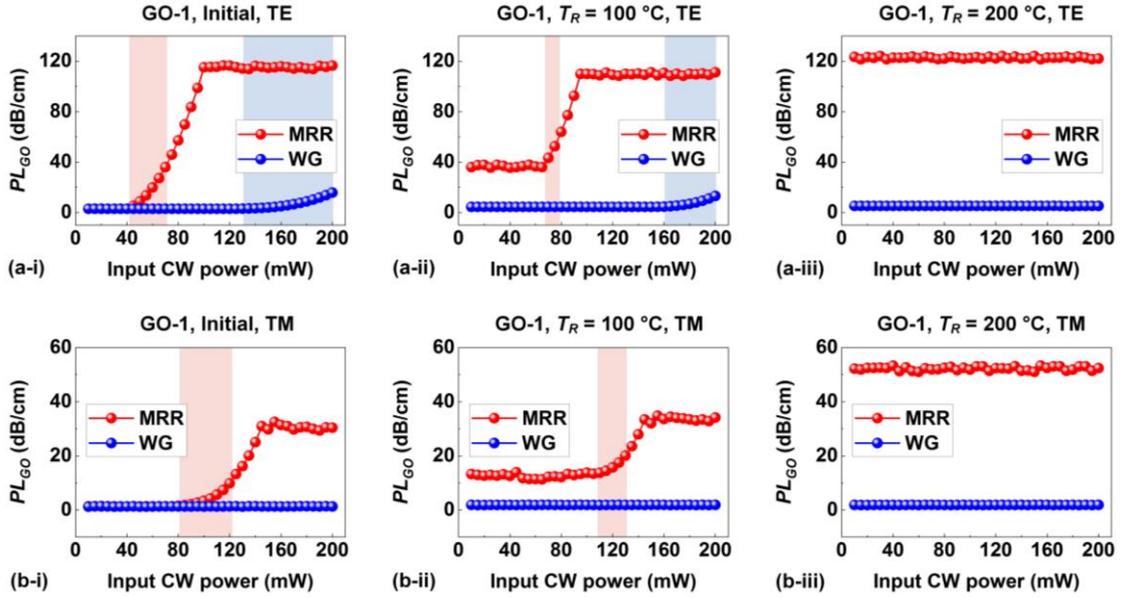

**Figure 7.** GO-induced propagation loss $PL_{GO}$ versus input CW power for a GO-coated SiN waveguide (WG) and a GO-coated SiN MRR for (**a**) TE and (**b**) TM polarizations. Both the WG and the MRR were coated with 1 layer of GO. In (**a**) – (**b**), (**i**) – (**iii**) show the results measured for a chip before heating (initial) and after the chip was heated at $T_R$ = ~100 °C and $T_R$ = ~200 °C, respectively. The blue and red shaded areas indicate the power ranges associated with reversible GO reduction for the WGs and the MRRs, respectively.

In **Figure 7**, we compare the reversible GO reduction behaviours in GO-coated SiN MRRs and waveguides. In our measurements, the GO-coated MRRs had a radius of ~200 μm, and the GO film length was ~50 μm (*i.e.*, the same as those in **Figures 5** and **6**). On the other hand, the GO-coated waveguides had a total length of ~2 cm, and the GO film length was ~1.4 mm. Since the MRRs and the waveguides had different GO film lengths, in **Figure 7** we compare the additional propagation loss induced by GO ($PL_{GO}$, in unit of dB/cm). For the hybrid MRRs, the $PL_{GO}$ was calculated by using the scattering matrix method to fit the measured transmission spectra. For the hybrid waveguides, the $PL_{GO}$ was calculated by dividing the GO-induced insertion loss by the GO film length. The $PL_{GO}$ in **Figure 7** was plotted as a function of input power. For GO-coated MRRs, the input power corresponds to the CW pump power $P_P$ injected into the MRR's resonance (similar to those in **Figures 5** and **6**). Whereas for the GO-coated waveguides, the input power corresponds to the input CW power coupled into the waveguide.

**Figure 7a-i** shows $PL_{GO}$ versus input CW power for a hybrid MRR and waveguide, both coated with 1 layer of GO. For the hybrid MRR, the trend of $PL_{GO}$ varying with input power matches the trend of $ER$ varying with $P_P$ in **Figure 5a-i**, with reversible GO reduction being observed within the same input power range of ~40 – 72 mW. The $PL_{GO}$ remained constant at ~3.0 dB/cm for a low input power < 40 mW, and at an input power of ~100 mW it dramatically increased to ~115.2 dB/cm. On the other hand, the $PL_{GO}$ for the hybrid waveguide remained constant at ~3.0 dB/cm for an input power < 120 mW, followed by a gradual increase to ~15.8 dB/cm when the input power reached ~200 mW. In the input power range of ~131 – 200 mW, reversible GO reduction behaviour was also observed. The comparison between the results for the hybrid MRR and waveguide reveals that there were more significant photo-thermal effects in the MRR due to resonant enhancement of optical intensity, which induced more significant changes to the GO film loss.

**Figures 7a-ii** and **7a-iii** show the corresponding results for the hybrid devices with 1 layer of reduced GO at $T_R$ = ~100 °C and ~200 °C, respectively. For both the hybrid MRR and waveguide, the input power range for reversible GO reduction narrows as the reduction degree increases – similar to the trend observed in **Figure 6**. At $T_R$ = ~200 °C, no obvious GO reduction was observed for either the MRR or the waveguide, as evidenced by the absence of significant changes in the $PL_{GO}$. **Figure 7b** shows the corresponding results for TM polarization. Similar to that in **Figures 5** and **6**, the TM-polarized hybrid MRRs showed higher thresholds and larger power ranges for reversible GO reduction than comparable TE-polarized MRRs. On the other hand, all the hybrid waveguides in TM polarization exhibited no significant changes in $PL_{GO}$ within our measured input power range, reflecting that there was no significant GO reduction even at a high input power of ~200 mW.

## 7. Optical bistability

Optical bistability featured by a steepened asymmetric transitional edge is a phenomenon arising from nonlinear thermo-optic effects. It has found wide applications in optical switches, memories, and logic devices [55-57]. In this section, we characterize the optical bistability in the fabricated GO-SiN MRRs. For the experiments performed in this section, we employed SiN MRRs with a radius of ~200 μm, and the length of the coated GO films was ~50 μm. We measured the output power of a CW light with gradually varying input power. The initial wavelength detuning $\delta$ between the CW light and the resonance wavelength of the MRR was defined as:

$$\delta = (\lambda_{laser} - \lambda_{res})/\Delta\lambda \tag{5}$$

where $\lambda_{laser}$ is the wavelength of the input CW light, $\lambda_{res}$ is the resonance wavelength of the MRR measured at a low input power of $P_{in} = \sim1$ mW (*i.e.*, the same as that in **Figure 2a**), and $\Delta\lambda$ is the full width at half-maximum (FWHM) of the resonance. For comparison, we selected the same $\delta = 0.2$ for all the measurements in this section.

**Figures 8a** and **b** show the measured output power $P_{out}$ versus input power $P_{in}$ for TE- and TM- polarized resonances of the uncoated SiN MRR, respectively. We first performed upward sweeping by gradually increasing $P_{in}$ from ~1 mW to ~28 mW, followed by downward sweeping where $P_{in}$ was gradually decreased within the same range. As can be seen, for both polarizations, the output power first gradually increased before experiencing a sudden drop toward a lower output power during the upward sweeping. In contrast, during the downward sweeping the output power first exhibited a gradual decrease before a sudden jump toward a higher output power. The existence of a hysteresis loop formed by the upward and downward sweeping confirms the presence of optical bistability in the SiN MRR. In principle, optical bistability manifests in the response of resonators when, under certain conditions, the output power produces multiple distinct solutions for a given input power [58]. As a result, the resonator can transition between the states corresponding to different solutions due to the influence of noise.

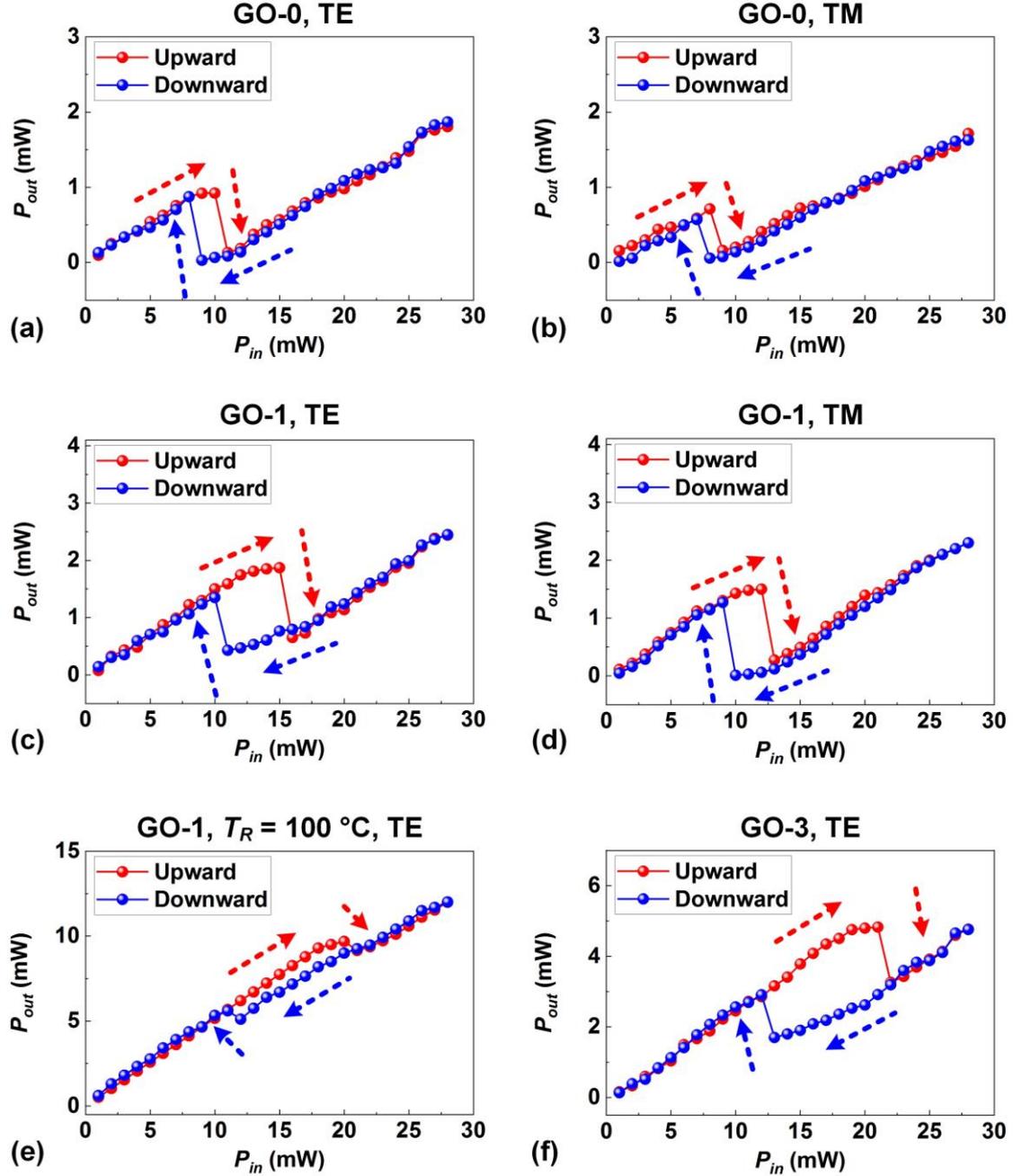

**Figure 8.** (**a**) – (**b**) Measured output power $P_{out}$ versus input CW power $P_{in}$ for TE- and TM-polarized resonances of an uncoated SiN MRR (GO-0), respectively. (**c**) – (**d**) Measured $P_{out}$ versus $P_{in}$ for TE- and TM-polarized resonances of a hybrid MRR coated with 1 layer of GO (GO-1), respectively. (**e**) Measured $P_{out}$ versus $P_{in}$ for a TE-polarized resonance of a hybrid MRR coated with 1 layer of GO after the chip was heated at $T_R = \sim100$ °C. (**f**) Measured $P_{out}$ versus $P_{in}$ for a TE-polarized resonance of a hybrid MRR coated with 3 layers of GO. In (**a**) – (**f**), the red and blue arrows indicate the increasing and decreasing of the input power, respectively. The initial wavelength detuning was $\delta = 0.2$.

**Figures 8c** and **d** show the measured output power $P_{out}$ versus input power $P_{in}$ for TE- and TM-polarized resonances of a hybrid MRR with 1 layer of GO, respectively. The power range for the upward and downward sweeping was the same as those in **Figures 8a** and **b**. Compared to the uncoated SiN MRR, the GO-coated MRR exhibited a more open hysteresis loop as well as lower power thresholds for the jump, reflecting the enhanced optical bistability behavior in the hybrid MRR. Such enhancement can be attributed to more significant thermo-optic effects within the GO films, as confirmed by the experimental results in previous sections. Unlike similar hysteresis loops that observed for TE- and TM-polarized resonances of the uncoated MRR, the hybrid MRR showed a more open hysteresis loop for the TE-polarized resonance compared to the TM-polarized resonance. This further confirms that the enhanced optical bistability was induced by the 2D GO film that had significant material anisotropy.

**Figures 8e** and **f** show the corresponding results for the TE-polarized resonances of a hybrid MRR with 1 layer of reduced GO (at $T_R = $ ~100 °C) and a hybrid MRR with 3 layers of GO, respectively. We did not use the hybrid MRR heated at $T_R = $ ~200 °C due to its low $ER$, which prevented us from observing obvious power jumps in the hysteresis loop. Compared to the hybrid MRR with unreduced GO, the hybrid MRR with reduced GO exhibited a broader input power range and a smaller output power range within the hysteresis loop. The former indicates higher thermal nonlinearity for reduced GO film, and the latter was mainly resulting from the decreased $ER$. Similar trends in the hysteresis loop were also observed for the hybrid MRR with a thicker GO film, albeit with a minor reduction in the output power range due to a smaller decrease in the $ER$. This reflects the increased thermal nonlinearity for thicker GO films. These results have wide applications to GO based devices as well as other novel photonic platforms. [59-96] Ultimately this could be useful for both classical and quantum microcomb based applications. [97-169]

## 8. Conclusion

In summary, we systematically investigate a series of thermo-optic properties of 2D layered GO films by precisely integrating them onto SiN MRRs. We characterize the refractive index, extinction coefficient, thermo-optic coefficient, and thermal conductivity of 2D layered GO films with different layer numbers and reduction degrees. Experimental results show broad variation ranges for their thermo-optic properties with increasing reduction degree, including an increase of ~0.2280 in the refractive index, a ~36 times improvement in the extinction coefficient, a transition from a positive thermo-optic coefficient to a negative one, and an over 48-fold increase in the thermal conductivity. In addition, the 2D GO films exhibit significant anisotropic response for light in TE and TM polarizations, including a difference of ~0.1400 for the refractive index, a ratio of ~4 for the extinction ratio, a ratio of ~6 for the thermo-optic coefficient, and a ratio of ~18 for the thermal conductivity. Finally, we demonstrate reversible GO reduction and enhanced optical bistability in the hybrid MRRs induced by photo-thermal effects. These results reveal the versatile thermo-optic properties of 2D GO, which greatly enrich potential functionalities and devices that can be developed for a range of thermo-optic applications.

## Conflict of interest

The authors declare no competing financial interest.

## References

[1]     T. L. Bergman, A. S. Lavine, F. P. Incropera, and D. P. DeWitt, *Introduction to heat transfer*. John Wiley & Sons, 2011.

[2]     D. M. Bierman, A. Lenert, W. R. Chan, B. Bhatia, I. Celanović, M. Soljačić, and E. N. Wang, "Enhanced photovoltaic energy conversion using thermally based spectral shaping," *Nature Energy,* vol. 1, no. 6, pp. 16068, 2016/05/23, 2016.

[3]     M. L. Brongersma, Y. Cui, and S. Fan, "Light management for photovoltaics using high-index nanostructures," *Nature Materials,* vol. 13, no. 5, pp. 451-460, 2014/05/01, 2014.

[4]     D. Li, X. Liu, W. Li, Z. Lin, B. Zhu, Z. Li, J. Li, B. Li, S. Fan, J. Xie, and J. Zhu, "Scalable and hierarchically designed polymer film as a selective thermal emitter for high-performance all-day radiative cooling," *Nature Nanotechnology,* vol. 16, no. 2, pp. 153-158,

2021/02/01, 2021.

[5]     L. Peng, D. Liu, H. Cheng, S. Zhou, and M. Zu, "A Multilayer Film Based Selective Thermal Emitter for Infrared Stealth Technology," *Advanced Optical Materials,* vol. 6, no. 23, pp. 1801006, 2018.

[6]     X. Xue, Y. Xuan, Y. Liu, P.-H. Wang, S. Chen, J. Wang, D. E. Leaird, M. Qi, and A. M. Weiner, "Mode-locked dark pulse Kerr combs in normal-dispersion microresonators," *Nature Photonics,* vol. 9, no. 9, pp. 594-600, 2015/09/01, 2015.

[7]     M. Rowley, P.-H. Hanzard, A. Cutrona, H. Bao, S. T. Chu, B. E. Little, R. Morandotti, D. J. Moss, G.-L. Oppo, J. S. Totero Gongora, M. Peccianti, and A. Pasquazi, "Self-emergence of robust solitons in a microcavity," *Nature,* vol. 608, no. 7922, pp. 303-309, 2022/08/01, 2022.

[8]     M. R. Watts, J. Sun, C. DeRose, D. C. Trotter, R. W. Young, and G. N. Nielson, "Adiabatic thermo-optic Mach-Zehnder switch," *Optics Letters,* vol. 38, no. 5, pp. 733-735, 2013/03/01, 2013.

[9]     A. Densmore, S. Janz, R. Ma, J. H. Schmid, D.-X. Xu, A. Delâge, J. Lapointe, M. Vachon, and P. Cheben, "Compact and low power thermo-optic switch using folded silicon waveguides," *Optics Express,* vol. 17, no. 13, pp. 10457-10465, 2009/06/22, 2009.

[10]    L. Wang, and B. Li, "Thermal Logic Gates: Computation with Phonons," *Physical Review Letters,* vol. 99, no. 17, pp. 177208, 10/24/, 2007.

[11]    S. Pal, and I. K. Puri, "Thermal AND Gate Using a Monolayer Graphene Nanoribbon," *Small,* vol. 11, no. 24, pp. 2910-2917, 2015.

[12]    C. Wan, Z. Zhang, J. Salman, J. King, Y. Xiao, Z. Yu, A. Shahsafi, R. Wambold, S. Ramanathan, and M. A. Kats, "Ultrathin Broadband Reflective Optical Limiter," *Laser & Photonics Reviews,* vol. 15, no. 6, pp. 2100001, 2021.

[13]    J. King, C. Wan, T. J. Park, S. Deshpande, Z. Zhang, S. Ramanathan, and M. A. Kats, "Electrically tunable VO2–metal metasurface for mid-infrared switching, limiting and nonlinear isolation," *Nature Photonics,* vol. 18, no. 1, pp. 74-80, 2024/01/01, 2024.

[14]    O. Kraieva, C. M. Quintero, I. Suleimanov, E. Hernandez, M, D. Lagrange, L. Salmon, W. Nicolazzi, G. Molnár, C. Bergaud, and A. Bousseksou, "High Spatial Resolution Imaging of Transient Thermal Events Using Materials with Thermal Memory," *Small,* vol. 12, no. 46, pp. pp.6325-6331, 2016-12-01, 2016.

[15]    A. M. Morsy, R. Biswas, and M. L. Povinelli, "High temperature, experimental thermal memory based on optical resonances in photonic crystal slabs," *APL Photonics,* vol. 4, no. 1, 2019.

[16]    F. N. Hamada, M. Rosenzweig, K. Kang, S. R. Pulver, A. Ghezzi, T. J. Jegla, and P. A. Garrity, "An internal thermal sensor controlling temperature preference in Drosophila," *Nature,* vol. 454, no. 7201, pp. 217-220, 2008/07/01, 2008.

[17]    Q. Hu, K.-T. Lin, H. Lin, Y. Zhang, and B. Jia, "Graphene Metapixels for Dynamically Switchable Structural Color," *ACS Nano,* vol. 15, no. 5, pp. 8930-8939, 2021/05/25, 2021.

[18]    Y. Li, W. Li, T. Han, X. Zheng, J. Li, B. Li, S. Fan, and C.-W. Qiu, "Transforming heat transfer with thermal metamaterials and devices," *Nature Reviews Materials,* vol. 6, no. 6, pp. 488-507, 2021/06/01, 2021.

[19]    J. Guo, G. Xu, D. Tian, Z. Qu, and C.-W. Qiu, "A Real-Time Self-Adaptive Thermal Metasurface," *Advanced Materials,* vol. 34, no. 24, pp. 2201093, 2022.

[20]    X. Gu, Y. Wei, X. Yin, B. Li, and R. Yang, "Colloquium: Phononic thermal properties of two-dimensional materials," *Reviews of Modern Physics,* vol. 90, no. 4, pp. 041002, 11/13/, 2018.

[21]    Y. Wang, N. Xu, D. Li, and J. Zhu, "Thermal Properties of Two Dimensional Layered Materials," *Advanced Functional Materials,* vol. 27, no. 19, pp. 1604134, 2017.

[22]    F. Wu, H. Tian, Y. Shen, Z.-Q. Zhu, Y. Liu, T. Hirtz, R. Wu, G. Gou, Y. Qiao, Y. Yang, C.-Y. Xing, G. Zhang, and T.-L. Ren, "High Thermal Conductivity 2D Materials: From Theory and Engineering to Applications," *Advanced Materials Interfaces,* vol. 9, no. 21, pp. 2200409, 2022.

[23]    H. Song, J. Liu, B. Liu, J. Wu, H.-M. Cheng, and F. Kang, "Two-Dimensional Materials for Thermal Management Applications," *Joule,* vol. 2, no. 3, pp. 442-463, 2018/03/21/, 2018.

[24]    J. Wu, H. Lin, D. J. Moss, K. P. Loh, and B. Jia, "Graphene oxide for photonics, electronics and optoelectronics," *Nature Reviews Chemistry,* vol. 7, no. 3, pp. 162-183, 2023/03/01, 2023.

[25]    J. Wu, L. Jia, Y. Zhang, Y. Qu, B. Jia, and D. J. Moss, "Graphene Oxide for Integrated Photonics and Flat Optics," *Advanced Materials,* vol. 33, no. 3, pp. 2006415, 2021.

[26]    K. P. Loh, Q. Bao, G. Eda, and M. Chhowalla, "Graphene oxide as a chemically tunable platform for optical applications," *Nature Chemistry,* vol. 2, no. 12, pp. 1015-1024, 2010/12/01, 2010.

[27]    H. Lin, B. C. Sturmberg, K.-T. Lin, Y. Yang, X. Zheng, T. K. Chong, C. M. de Sterke, and B. Jia, "A 90-nm-thick graphene metamaterial for strong and extremely broadband absorption of unpolarized light," *Nature Photonics,* vol. 13, no. 4, pp. 270-276, 2019.

[28]    J. Wu, Y. Yang, Y. Qu, X. Xu, Y. Liang, S. T. Chu, B. E. Little, R. Morandotti, B. Jia, and D. J. Moss, "Graphene Oxide Waveguide and Micro-Ring Resonator Polarizers," *Laser & Photonics Reviews,* vol. 13, no. 9, pp. 1900056, 2019.

[29]    Y. Qu, J. Wu, Y. Zhang, Y. Yang, J. Linnan, H. el Dirani, S. Kerdiles, C. Sciancalepore, P. Demongodin, C. Grillet, C. Monat, B. Jia, and D. Moss, "Integrated optical parametric amplifiers in silicon nitride waveguides incorporated with 2D graphene oxide films," *Light: Advanced Manufacturing,* vol. 4, pp. 1, 01/01, 2023.

[30]    X. Zheng, B. Jia, H. Lin, L. Qiu, D. Li, and M. Gu, "Highly efficient and ultra-broadband graphene oxide ultrathin lenses with three-dimensional subwavelength focusing," *Nature Communications,* vol. 6, no. 1, pp. 8433, 2015/09/22, 2015.

[31]    S. Wei, G. Cao, H. Lin, X. Yuan, M. Somekh, and B. Jia, "A Varifocal Graphene Metalens for Broadband Zoom Imaging Covering the Entire Visible Region," *ACS Nano,* vol. 15, no. 3, pp. 4769-4776, 2021/03/23, 2021.

[32]    Y. Yang, Y. Zhang, J. Zhang, X. Zheng, Z. Gan, H. Lin, M. Hong, and B. Jia, "Graphene Metamaterial 3D Conformal Coating for Enhanced Light Harvesting," *ACS Nano,* vol. 17, no. 3, pp. 2611-2619, 2023/02/14, 2023.

[33]    J. Wu, Y. Yang, Y. Qu, L. Jia, Y. Zhang, X. Xu, S. T. Chu, B. E. Little, R. Morandotti, B. Jia, and D. J. Moss, "2D Layered Graphene Oxide Films Integrated with Micro-Ring Resonators for Enhanced Nonlinear Optics," *Small,* vol. 16, no. 16, pp. 1906563, 2020.

[34]    K.-T. Lin, H. Lin, T. Yang, and B. Jia, "Structured graphene metamaterial selective absorbers for high efficiency and omnidirectional solar thermal energy conversion," *Nature Communications,* vol. 11, no. 1, pp. 1389, 2020/03/13, 2020.

[35]     Z. Luo, P. M. Vora, E. J. Mele, A. T. C. Johnson, and J. M. Kikkawa, "Photoluminescence and band gap modulation in graphene oxide," *Applied Physics Letters,* vol. 94, no. 11, 2009.

[36]     Y. Zhang, J. Wu, L. Jia, Y. Qu, Y. Yang, B. Jia, and D. J. Moss, "Graphene Oxide for Nonlinear Integrated Photonics," *Laser & Photonics Reviews,* vol. 17, no. 3, pp. 2200512, 2023/03/01, 2023.

[37]     A. M. Dimiev, and S. Eigler, *Graphene oxide: fundamentals and applications*: John Wiley & Sons, 2016.

[38]     H. El Dirani, L. Youssef, C. Petit-Etienne, S. Kerdiles, P. Grosse, C. Monat, E. Pargon, and C. Sciancalepore, "Ultralow-loss tightly confining Si3N4 waveguides and high-Q microresonators," *Optics Express,* vol. 27, no. 21, pp. 30726-30740, 2019/10/14, 2019.

[39]     Y. Yang, H. Lin, B. Y. Zhang, Y. Zhang, X. Zheng, A. Yu, M. Hong, and B. Jia, "Graphene-based multilayered metamaterials with phototunable architecture for on-chip photonic devices," *Acs Photonics,* vol. 6, no. 4, pp. 1033-1040, 2019.

[40]     H. Arianfard, S. Juodkazis, D. J. Moss, and J. Wu, "Sagnac interference in integrated photonics," *Applied Physics Reviews,* vol. 10, no. 1, 2023.

[41]     R. M. Vázquez, S. M. Eaton, R. Ramponi, G. Cerullo, and R. Osellame, "Fabrication of binary Fresnel lenses in PMMA by femtosecond laser surface ablation," *Optics Express,* vol. 19, no. 12, pp. 11597-11604, 2011/06/06, 2011.

[42]     X. Li, H. Ren, X. Chen, J. Liu, Q. Li, C. Li, G. Xue, J. Jia, L. Cao, A. Sahu, B. Hu, Y. Wang, G. Jin, and M. Gu, "Athermally photoreduced graphene oxides for three-dimensional holographic images," *Nature Communications,* vol. 6, no. 1, pp. 6984, 2015/04/22, 2015.

[43]     T.-Z. Shen, S.-H. Hong, and J.-K. Song, "Electro-optical switching of graphene oxide liquid crystals with an extremely large Kerr coefficient," *Nature Materials,* vol. 13, no. 4, pp. 394-399, 2014/04/01, 2014.

[44]     B. Frey, D. Leviton, and T. Madison, *Temperature-dependent refractive index of silicon and germanium*, p.^pp. AS: SPIE, 2006.

[45]     B. Guha, J. Cardenas, and M. Lipson, "Athermal silicon microring resonators with titanium oxide cladding," *Optics Express,* vol. 21, no. 22, pp. 26557-26563, 2013/11/04, 2013.

[46]     A. Arbabi, and L. L. Goddard, "Measurements of the refractive indices and thermo-optic coefficients of Si3N4 and SiOx using microring resonances," *Optics Letters,* vol. 38, no. 19, pp. 3878-3881, 2013/10/01, 2013.

[47]     C. Horvath, D. Bachman, R. Indoe, and V. Van, "Photothermal nonlinearity and optical bistability in a graphene–silicon waveguide resonator," *Optics Letters,* vol. 38, no. 23, pp. 5036-5039, 2013.

[48]     W. Bogaerts, P. De Heyn, T. Van Vaerenbergh, K. De Vos, S. Kumar Selvaraja, T. Claes, P. Dumon, P. Bienstman, D. Van Thourhout, and R. Baets, "Silicon microring resonators," *Laser & Photonics Reviews,* vol. 6, no. 1, pp. 47-73, 2012/01/02, 2012.

[49]     Y. Gao, W. Zhou, X. Sun, H. K. Tsang, and C. Shu, "Cavity-enhanced thermo-optic bistability and hysteresis in a graphene-on-Si3N4 ring resonator," *Optics Letters,* vol. 42, no. 10, pp. 1950-1953, 2017/05/15, 2017.

[50]     Y. Li, H. Lin, and N. Mehra, "Identification of Thermal Barrier Areas in Graphene Oxide/Boron Nitride Membranes by Scanning Thermal Microscopy: Thermal Conductivity Improvement through Membrane Assembling," *ACS Applied Nano Materials,* vol. 4, no. 4, pp. 4189-4198, 2021/04/23, 2021.

[51]     A. A. Balandin, S. Ghosh, W. Bao, I. Calizo, D. Teweldebrhan, F. Miao, and C. N. Lau, "Superior Thermal Conductivity of Single-Layer Graphene," *Nano Letters,* vol. 8, no. 3, pp. 902-907, 2008/03/01, 2008.

[52]     J. D. Renteria, S. Ramirez, H. Malekpour, B. Alonso, A. Centeno, A. Zurutuza, A. I. Cocemasov, D. L. Nika, and A. A. Balandin, "Strongly Anisotropic Thermal Conductivity of Free-Standing Reduced Graphene Oxide Films Annealed at High Temperature," *Advanced Functional Materials,* vol. 25, no. 29, pp. 4664-4672, 2015/08/01, 2015.

[53]     S. Nakamura, K. Sekiya, S. Matano, Y. Shimura, Y. Nakade, K. Nakagawa, Y. Monnai, and H. Maki, "High-Speed and On-Chip Optical Switch Based on a Graphene Microheater," *ACS Nano,* vol. 16, no. 2, pp. 2690-2698, 2022/02/22, 2022.

[54]     G.-J. Zhou, and W.-Y. Wong, "Organometallic acetylides of PtII, AuI and HgII as new generation optical power limiting materials," *Chemical Society Reviews,* vol. 40, no. 5, pp. 2541-2566, 2011.

[55]     P. W. Smith, and W. J. Tomlinson, "Bistable optical devices promise subpicosecond switching," *IEEE Spectrum,* vol. 18, pp. 26-33, June 01, 1981, 1981.

[56]     H. Gibbs, "Optical Bistability: Controlling Light with Light, Academic Press," *Inc.: Orlando, FL, USA,* 1985.

[57]     V. R. Almeida, and M. Lipson, "Optical bistability on a silicon chip," *Optics Letters,* vol. 29, no. 20, pp. 2387-2389, 2004/10/15, 2004.

[58]     A. A. Nikitin, I. A. Ryabcev, A. A. Nikitin, A. V. Kondrashov, A. A. Semenov, D. A. Konkin, A. A. Kokolov, F. I. Sheyerman, L. I. Babak, and A. B. Ustinov, "Optical bistable SOI micro-ring resonators for memory applications," *Optics Communications,* vol. 511, pp. 127929, 2022/05/15/, 2022.

59.     Wu, J. et al. 2D layered graphene oxide films integrated with micro-ring resonators for enhanced nonlinear optics. Small Vol. 16, 1906563 (2020).

60.     Wu, J. et al. Graphene oxide waveguide and micro-ring resonator polarizers. Laser Photonics Rev. Vol. 13, 1900056 (2019).

61.     Zhang, Y. et al. Enhanced Kerr nonlinearity and nonlinear figure of merit in silicon nanowires integrated with 2d graphene oxide films. ACS Appl. Mater. Interfaces Vol. 12, 33094-33103 (2020).

62.     Qu, Y. et al. Enhanced four-wave mixing in silicon nitride waveguides integrated with 2d layered graphene oxide films. Advanced Optical Materials Vol. 8, 2001048 (2020).

63.     Jia, L. et al. Highly nonlinear BiOBr nanoflakes for hybrid integrated photonics. APL Photonics Vol. 4, 090802 (2019).

64.     Yuning Zhang, Jiayang Wu, Yang Qu, Yunyi Yang, Linnan Jia, Baohua Jia, and David J. Moss, "Enhanced supercontinuum generated in SiN waveguides coated with GO films", Advanced Materials Technologies Vol. 8 (1) 2201796 (2023).

65.     Yuning Zhang, Jiayang Wu, Linnan Jia, Yang Qu, Baohua Jia, and David J. Moss, "Graphene oxide for nonlinear integrated photonics", Laser and Photonics Reviews Vol. 17 2200512 (2023). DOI:10.1002/lpor.202200512.

66. Jiayang Wu, H. Lin, D. J. Moss, T.K. Loh, Baohua Jia, "Graphene oxide: new opportunities for electronics, photonics, and optoelectronics", Nature Reviews Chemistry Vol. 7 (3) 162–183 (2023). DOI:10.1038/s41570-022-00458-7.

67. Yang Qu, Jiayang Wu, Yuning Zhang, Yunyi Yang, Linnan Jia, Baohua Jia, and David J. Moss, "Photo thermal tuning in GO-coated integrated waveguides", Micromachines Vol. 13 1194 (2022). doi.org/10.3390/mi13081194

68. Yuning Zhang, Jiayang Wu, Yunyi Yang, Yang Qu, Houssein El Dirani, Romain Crochemore, Corrado Sciancalepore, Pierre Demongodin, Christian Grillet, Christelle Monat, Baohua Jia, and David J. Moss, "Enhanced self-phase modulation in silicon nitride waveguides integrated with 2D graphene oxide films", IEEE Journal of Selected Topics in Quantum Electronics Vol. 29 (1) 5100413 (2023). DOI: 10.1109/JSTQE.2022.3177385

69. Yuning Zhang, Jiayang Wu, Yunyi Yang, Yang Qu, Linnan Jia, Baohua Jia, and David J. Moss, "Enhanced spectral broadening of femtosecond optical pulses in silicon nanowires integrated with 2D graphene oxide films", Micromachines Vol. 13 756 (2022). DOI:10.3390/mi13050756.

70. Linnan Jia, Jiayang Wu, Yuning Zhang, Yang Qu, Baohua Jia, Zhigang Chen, and David J. Moss, "Fabrication Technologies for the On-Chip Integration of 2D Materials", Small: Methods Vol. 6, 2101435 (2022). DOI:10.1002/smtd.202101435.

71. Yuning Zhang, Jiayang Wu, Yang Qu, Linnan Jia, Baohua Jia, and David J. Moss, "Design and optimization of four-wave mixing in microring resonators integrated with 2D graphene oxide films", Journal of Lightwave Technology Vol. 39 (20) 6553-6562 (2021). DOI:10.1109/JLT.2021.3101292.

72. Yuning Zhang, Jiayang Wu, Yang Qu, Linnan Jia, Baohua Jia, and David J. Moss, "Optimizing the Kerr nonlinear optical performance of silicon waveguides integrated with 2D graphene oxide films", Journal of Lightwave Technology Vol. 39 (14) 4671-4683 (2021). DOI: 10.1109/JLT.2021.3069733.

73. Yang Qu, Jiayang Wu, Yuning Zhang, Yao Liang, Baohua Jia, and David J. Moss, "Analysis of four-wave mixing in silicon nitride waveguides integrated with 2D layered graphene oxide films", Journal of Lightwave Technology Vol. 39 (9) 2902-2910 (2021). DOI: 10.1109/JLT.2021.3059721.

74. Jiayang Wu, Linnan Jia, Yuning Zhang, Yang Qu, Baohua Jia, and David J. Moss, "Graphene oxide: versatile films for flat optics to nonlinear photonic chips", Advanced Materials Vol. 33 (3) 2006415, pp.1-29 (2021). DOI:10.1002/adma.202006415.

75. Y. Qu, J. Wu, Y. Zhang, L. Jia, Y. Yang, X. Xu, S. T. Chu, B. E. Little, R. Morandotti, B. Jia, and D. J. Moss, "Graphene oxide for enhanced optical nonlinear performance in CMOS compatible integrated devices", Paper No. 11688-30, PW21O-OE109-36, 2D Photonic Materials and Devices IV, SPIE Photonics West, San Francisco CA March 6-11 (2021). doi.org/10.1117/12.2583978

76. Yang Qu, Jiayang Wu, Yunyi Yang, Yuning Zhang, Yao Liang, Houssein El Dirani, Romain Crochemore, Pierre Demongodin, Corrado Sciancalepore, Christian Grillet, Christelle Monat, Baohua Jia, and David J. Moss, "Enhanced nonlinear four-wave mixing in silicon nitride waveguides integrated with 2D layered graphene oxide films", Advanced Optical Materials vol. 8 (21) 2001048 (2020). DOI: 10.1002/adom.202001048. arXiv:2006.14944.

77. Yuning Zhang, Yang Qu, Jiayang Wu, Linnan Jia, Yunyi Yang, Xingyuan Xu, Baohua Jia, and David J. Moss, "Enhanced Kerr nonlinearity and nonlinear figure of merit in silicon nanowires integrated with 2D graphene oxide films", ACS Applied Materials and Interfaces vol. 12 (29) 33094−33103 June 29 (2020). DOI:10.1021/acsami.0c07852

78. Jiayang Wu, Yunyi Yang, Yang Qu, Yuning Zhang, Linnan Jia, Xingyuan Xu, Sai T. Chu, Brent E. Little, Roberto Morandotti, Baohua Jia, and David J. Moss, "Enhanced nonlinear four-wave mixing in microring resonators integrated with layered graphene oxide films", Small vol. 16 (16) 1906563 (2020). DOI: 10.1002/smll.201906563

79. Jiayang Wu, Yunyi Yang, Yang Qu, Xingyuan Xu, Yao Liang, Sai T. Chu, Brent E. Little, Roberto Morandotti, Baohua Jia, and David J. Moss, "Graphene oxide waveguide polarizers and polarization selective micro-ring resonators", Paper 11282-29, SPIE Photonics West, San Francisco, CA, 4 - 7 February (2020). doi: 10.1117/12.2544584

80. Jiayang Wu, Yunyi Yang, Yang Qu, Xingyuan Xu, Yao Liang, Sai T. Chu, Brent E. Little, Roberto Morandotti, Baohua Jia, and David J. Moss, "Graphene oxide waveguide polarizers and polarization selective micro-ring resonators", Laser and Photonics Reviews vol. 13 (9) 1900056 (2019). DOI:10.1002/lpor.201900056.

81. Yunyi Yang, Jiayang Wu, Xingyuan Xu, Sai T. Chu, Brent E. Little, Roberto Morandotti, Baohua Jia, and David J. Moss, "Enhanced four-wave mixing in graphene oxide coated waveguides", Applied Physics Letters Photonics vol. 3 120803 (2018). doi: 10.1063/1.5045509.

82. Linnan Jia, Yang Qu, Jiayang Wu, Yuning Zhang, Yunyi Yang, Baohua Jia, and David J. Moss, "Third-order optical nonlinearities of 2D materials at telecommunications wavelengths", Micromachines (MDPI), 14, 307 (2023). https://doi.org/10.3390/mi14020307.

83. Linnan Jia, Dandan Cui, Jiayang Wu, Haifeng Feng, Tieshan Yang, Yunyi Yang, Yi Du, Weichang Hao, Baohua Jia, David J. Moss, "BiOBr nanoflakes with strong nonlinear optical properties towards hybrid integrated photonic devices", Applied Physics Letters Photonics vol. 4 090802 vol. (2019). DOI: 10.1063/1.5116621

84. Linnan Jia, Jiayang Wu, Yunyi Yang, Yi Du, Baohua Jia, David J. Moss, "Large Third-Order Optical Kerr Nonlinearity in Nanometer-Thick PdSe2 2D Dichalcogenide Films: Implications for Nonlinear Photonic Devices", ACS Applied Nano Materials vol. 3 (7) 6876–6883 (2020). DOI:10.1021/acsanm.0c01239.

85. A. Pasquazi, et al., "Sub-picosecond phase-sensitive optical pulse characterization on a chip", Nature Photonics, vol. 5, no. 10, pp. 618-623 (2011).

86. M Ferrera et al., "On-Chip ultra-fast 1st and 2nd order CMOS compatible all-optical integration", Optics Express vol. 19 (23), 23153-23161 (2011).

87. L Carletti, P Ma, Y Yu, B Luther-Davies, D Hudson, C Monat, …. , et al., "Nonlinear optical response of low loss silicon germanium waveguides in the mid-infrared", Optics Express vol. 23 (7), 8261-8271 (2015).

88. Hamed Arianfard, Saulius Juodkazis, David J. Moss, and Jiayang Wu, "Sagnac interference in integrated photonics", Applied Physics Reviews vol. 10 (1) 011309 (2023). doi: 10.1063/5.0123236. (2023).

89. Hamed Arianfard, Jiayang Wu, Saulius Juodkazis, and David J. Moss, "Optical analogs of Rabi splitting in integrated waveguide-coupled resonators", Advanced Physics Research   2 (2023). DOI: 10.1002/apxr.202200123.

90. Hamed Arianfard, Jiayang Wu, Saulius Juodkazis, and David J. Moss, "Spectral shaping based on optical waveguides with advanced Sagnac loop reflectors", Paper No. PW22O-OE201-20, SPIE-Opto, Integrated Optics: Devices, Materials, and Technologies XXVI, SPIE Photonics West, San Francisco CA January 22 - 27 (2022). doi: 10.1117/12.2607902

91. Hamed Arianfard, Jiayang Wu, Saulius Juodkazis, David J. Moss, "Spectral Shaping Based on Integrated Coupled Sagnac Loop Reflectors Formed by a Self-Coupled Wire Waveguide", IEEE Photonics Technology Letters vol. 33 (13) 680-683 (2021). DOI:10.1109/LPT.2021.3088089.

92. Hamed Arianfard, Jiayang Wu, Saulius Juodkazis and David J. Moss, "Three Waveguide Coupled Sagnac Loop Reflectors for Advanced Spectral Engineering", Journal of Lightwave Technology vol. 39 (11) 3478-3487 (2021). DOI: 10.1109/JLT.2021.3066256.

93. Hamed Arianfard, Jiayang Wu, Saulius Juodkazis and David J. Moss, "Advanced Multi-Functional Integrated Photonic Filters based on Coupled Sagnac Loop Reflectors", Journal of Lightwave Technology vol. 39 Issue: 5, pp.1400-1408 (2021). DOI:10.1109/JLT.2020.3037559.

94. Hamed Arianfard, Jiayang Wu, Saulius Juodkazis and David J. Moss, "Advanced multi-functional integrated photonic filters based on coupled Sagnac loop reflectors", Paper 11691-4, PW21O-OE203-44, Silicon Photonics XVI, SPIE Photonics West, San Francisco CA March 6-11 (2021).   doi.org/10.1117/12.2584020

95. Jiayang Wu, Tania Moein, Xingyuan Xu, and David J. Moss, "Advanced photonic filters via cascaded Sagnac loop reflector resonators in silicon-on-insulator integrated nanowires", Applied Physics Letters Photonics vol. 3 046102 (2018). DOI:/10.1063/1.5025833

96. Jiayang Wu, Tania Moein, Xingyuan Xu, Guanghui Ren, Arnan Mitchell, and David J. Moss, "Micro-ring resonator quality factor enhancement via an integrated Fabry-

Perot cavity", Applied Physics Letters Photonics vol. 2 056103 (2017). doi: 10.1063/1.4981392.

97.  M Ferrera, Y Park, L Razzari, BE Little, ST Chu, R Morandotti, DJ Moss, ... et al., "All-optical 1st and 2nd order integration on a chip", Optics Express vol. 19 (23), 23153-23161 (2011).

98.  Bao, C., et al., Direct soliton generation in microresonators, Opt. Lett, 42, 2519 (2017).

99.  M.Ferrera et al., "CMOS compatible integrated all-optical RF spectrum analyzer", Optics Express, vol. 22, no. 18, 21488 - 21498 (2014).

100. M. Kues, et al., "Passively modelocked laser with an ultra-narrow spectral width", Nature Photonics, vol. 11, no. 3, pp. 159, 2017.

101. M.Ferrera et al."On-Chip ultra-fast 1st and 2nd order CMOS compatible all-optical integration", Opt. Express, vol. 19, (23)pp. 23153-23161 (2011).

102. D. Duchesne, M. Peccianti, M. R. E. Lamont, et al., "Supercontinuum generation in a high index doped silica glass spiral waveguide," Optics Express, vol. 18, no, 2, pp. 923-930, 2010.

103. H Bao, L Olivieri, M Rowley, ST Chu, BE Little, R Morandotti, DJ Moss, ... "Turing patterns in a fiber laser with a nested microresonator: Robust and controllable microcomb generation", Physical Review Research vol. 2 (2), 023395 (2020).

104. M. Ferrera, et al., "On-chip CMOS-compatible all-optical integrator", Nature Communications, vol. 1, Article 29, 2010.

105. A. Pasquazi, et al., "All-optical wavelength conversion in an integrated ring resonator," Optics Express, vol. 18, no. 4, pp. 3858-3863, 2010.

106. A.Pasquazi, Y. Park, J. Azana, et al., "Efficient wavelength conversion and net parametric gain via Four Wave Mixing in a high index doped silica waveguide," Optics Express, vol. 18, no. 8, pp. 7634-7641, 2010.

107. Peccianti, M. Ferrera, L. Razzari, et al., "Subpicosecond optical pulse compression via an integrated nonlinear chirper," Optics Express, vol. 18, no. 8, pp. 7625-7633, 2010.

108. M. Ferrera et al., "Low Power CW Parametric Mixing in a Low Dispersion High Index Doped Silica Glass Micro-Ring Resonator with Q-factor > 1 Million", Optics Express, vol.17, no. 16, pp. 14098–14103 (2009).

109. M. Peccianti, et al., "Demonstration of an ultrafast nonlinear microcavity modelocked laser", Nature Communications, vol. 3, pp. 765, 2012.

110. A.Pasquazi, et al., "Self-locked optical parametric oscillation in a CMOS compatible microring resonator: a route to robust optical frequency comb generation on a chip," Optics Express, vol. 21, no. 11, pp. 13333-13341, 2013.

111. A.Pasquazi, et al., "Stable, dual mode, high repetition rate mode-locked laser based on a microring resonator," Optics Express, vol. 20, no. 24, pp. 27355-27362, 2012.

112. Pasquazi, A. et al. Micro-combs: a novel generation of optical sources. Physics Reports 729, 1-81 (2018).

113. L. Razzari, et al., "CMOS-compatible integrated optical hyper-parametric oscillator," Nature Photonics, vol. 4, no. 1, pp. 41-45, 2010.

114. M. Ferrera, et al., "Low-power continuous-wave nonlinear optics in doped silica glass integrated waveguide structures," Nature Photonics, vol. 2, no. 12, pp. 737-740, 2008.

115. Moss, D. J., Morandotti, R., Gaeta, A. L. & Lipson, M. New CMOS compatible platforms based on silicon nitride and Hydex for nonlinear optics. Nat. Photonics Vol. 7, 597-607 (2013).

116. Sun, Y. et al. Applications of optical microcombs. Advances in Optics and Photonics Vol. 15, 86 (2023).

117. H. Bao, et al., Laser cavity-soliton microcombs, Nature Photonics, vol. 13, no. 6, pp. 384-389, Jun. 2019.

118. Antonio Cutrona, Maxwell Rowley, Debayan Das, Luana Olivieri, Luke Peters, Sai T. Chu, Brent L. Little, Roberto Morandotti, David J. Moss, Juan Sebastian Totero Gongora, Marco Peccianti, Alessia Pasquazi, "High Conversion Efficiency in Laser Cavity-Soliton Microcombs", Optics Express Vol. 30, Issue 22, pp. 39816-39825 (2022). https://doi.org/10.1364/OE.470376.

119. M.Rowley, P.Hanzard, A.Cutrona, H.Bao, S.Chu, B.Little, R.Morandotti, D. J. Moss, G. Oppo, J. Gongora, M. Peccianti and A. Pasquazi, "Self-emergence of robust solitons in a micro-cavity", Nature vol. 608 (7922) 303–309 (2022).

120. A. Cutrona, M. Rowley, A. Bendahmane, V. Cecconi,L. Peters, L. Olivieri, B. E. Little, S. T. Chu, S. Stivala, R. Morandotti, D. J. Moss, J. S. Totero-Gongora, M. Peccianti, A. Pasquazi, "Nonlocal bonding of a soliton and a blue-detuned state in a microcomb laser",   Nature Communications Physics vol. 6 (2023).

121. A. Cutrona, M. Rowley, A. Bendahmane, V. Cecconi,L. Peters, L. Olivieri, B. E. Little, S. T. Chu, S. Stivala, R. Morandotti, D. J. Moss, J. S. Totero-Gongora, M. Peccianti, A. Pasquazi, "Stability Properties of Laser Cavity-Solitons for Metrological Applications", Applied Physics Letters vol. 122 (12) 121104 (2023); doi: 10.1063/5.0134147.

122. X. Xu, J. Wu, M. Shoeiby, T. G. Nguyen, S. T. Chu, B. E. Little, R. Morandotti, A. Mitchell, and D. J. Moss, "Reconfigurable broadband microwave photonic intensity differentiator based on an integrated optical frequency comb source," APL Photonics, vol. 2, no. 9, 096104, Sep. 2017.

123. Xu, X., et al., Photonic microwave true time delays for phased array antennas using a 49 GHz FSR integrated micro-comb source, Photonics Research, vol. 6, B30-B36 (2018).

124. X. Xu, M. Tan, J. Wu, R. Morandotti, A. Mitchell, and D. J. Moss, "Microcomb-based photonic RF signal processing", IEEE Photonics Technology Letters, vol. 31 no. 23 1854-1857, 2019.

125. Xu, et al., "Advanced adaptive photonic RF filters with 80 taps based on an integrated optical micro-comb source," Journal of Lightwave Technology, vol. 37, no. 4, pp. 1288-1295 (2019).

126. X. Xu, et al., "Photonic RF and microwave integrator with soliton crystal microcombs", IEEE Transactions on Circuits and Systems II: Express Briefs, vol. 67, no. 12, pp. 3582-3586, 2020. DOI:10.1109/TCSII.2020.2995682.

127. X. Xu, et al., "High performance RF filters via bandwidth scaling with Kerr micro-combs," APL Photonics, vol. 4 (2) 026102. 2019.

128. M. Tan, et al., "Microwave and RF photonic fractional Hilbert transformer based on a 50 GHz Kerr micro-comb", Journal of Lightwave Technology, vol. 37, no. 24, pp. 6097 – 6104, 2019.

129. M. Tan, et al., "RF and microwave fractional differentiator based on photonics", IEEE Transactions on Circuits and Systems: Express Briefs, vol. 67, no.11, pp. 2767-2771, 2020. DOI:10.1109/TCSII.2020.2965158.

130. M. Tan, et al., "Photonic RF arbitrary waveform generator based on a soliton crystal micro-comb source", Journal of Lightwave Technology, vol. 38, no. 22, pp. 6221-6226 (2020). DOI: 10.1109/JLT.2020.3009655.

131. M. Tan, X. Xu, J. Wu, R. Morandotti, A. Mitchell, and D. J. Moss, "RF and microwave high bandwidth signal processing based on Kerr Micro-combs", Advances in Physics X, VOL. 6, NO. 1, 1838946 (2021). DOI:10.1080/23746149.2020.1838946.

132. X. Xu, et al., "Advanced RF and microwave functions based on an integrated optical frequency comb source," Opt. Express, vol. 26 (3) 2569 (2018).

133. M. Tan, X. Xu, J. Wu, B. Corcoran, A. Boes, T. G. Nguyen, S. T. Chu, B. E. Little, R.Morandotti, A. Lowery, A. Mitchell, and D. J. Moss, ""Highly Versatile Broadband RF Photonic Fractional Hilbert Transformer Based on a Kerr Soliton Crystal Microcomb", Journal of Lightwave Technology vol. 39 (24) 7581-7587 (2021).

134. Wu, J. et al. RF Photonics: An Optical Microcombs' Perspective. IEEE Journal of Selected Topics in Quantum Electronics Vol. 24, 6101020, 1-20 (2018).

135. T. G. Nguyen et al., "Integrated frequency comb source-based Hilbert transformer for wideband microwave photonic phase analysis," Opt. Express, vol. 23, no. 17, pp. 22087-22097, Aug. 2015.

136. X. Xu, et al., "Broadband RF channelizer based on an integrated optical frequency Kerr comb source," Journal of Lightwave Technology, vol. 36, no. 19, pp. 4519-4526, 2018.

137. X. Xu, et al., "Continuously tunable orthogonally polarized RF optical single sideband generator based on micro-ring resonators," Journal of Optics, vol. 20, no. 11, 115701. 2018.

138. X. Xu, et al., "Orthogonally polarized RF optical single sideband generation and dual-channel equalization based on an integrated microring resonator," Journal of Lightwave Technology, vol. 36, no. 20, pp. 4808-4818. 2018.

139. X. Xu, et al., "Photonic RF phase-encoded signal generation with a microcomb source", J. Lightwave Technology, vol. 38, no. 7, 1722-1727, 2020.

140. X. Xu, et al., Broadband microwave frequency conversion based on an integrated optical micro-comb source", Journal of Lightwave Technology, vol. 38 no. 2, pp. 332-338, 2020.

141. M. Tan, et al., "Photonic RF and microwave filters based on 49GHz and 200GHz Kerr microcombs", Optics Comm. vol. 465,125563, Feb. 22. 2020.

142. X. Xu, et al., "Broadband photonic RF channelizer with 90 channels based on a soliton crystal microcomb", Journal of Lightwave Technology, Vol. 38, no. 18, pp. 5116 – 5121 (2020). doi: 10.1109/JLT.2020.2997699.

143. M. Tan et al, "Orthogonally polarized Photonic Radio Frequency single sideband generation with integrated micro-ring resonators", IOP Journal of Semiconductors, Vol. 42 (4), 041305 (2021). DOI: 10.1088/1674-4926/42/4/041305.

144. Mengxi Tan, X. Xu, J. Wu, T. G. Nguyen, S. T. Chu, B. E. Little, R. Morandotti, A. Mitchell, and David J. Moss, "Photonic Radio Frequency Channelizers based on Kerr Optical Micro-combs", IOP Journal of Semiconductors Vol. 42 (4), 041302 (2021). DOI:10.1088/1674-4926/42/4/041302.

145. B. Corcoran, et al., "Ultra-dense optical data transmission over standard fiber with a single chip source", Nature Communications, vol. 11, Article:2568, 2020.

146. X. Xu et al, "Photonic perceptron based on a Kerr microcomb for scalable high speed optical neural networks", Laser and Photonics Reviews, vol. 14, no. 8, 2000070 (2020). DOI: 10.1002/lpor.202000070.

147. X. Xu, et al., "11 TOPs photonic convolutional accelerator for optical neural networks", Nature vol. 589, 44-51 (2021).

148. X. Xu et al., "Neuromorphic computing based on wavelength-division multiplexing", 28 IEEE Journal of Selected Topics in Quantum Electronics Vol. 29 Issue: 2, Article 7400112 (2023). DOI:10.1109/JSTQE.2022.3203159.

149. Yang Sun, Jiayang Wu, Mengxi Tan, Xingyuan Xu, Yang Li, Roberto Morandotti, Arnan Mitchell, and David Moss, "Applications of optical micro-combs", Advances in Optics and Photonics vol. 15 (1) 86-175 (2023). DOI:10.1364/AOP.470264.

150. Yunping Bai, Xingyuan Xu,1, Mengxi Tan, Yang Sun, Yang Li, Jiayang Wu, Roberto Morandotti, Arnan Mitchell, Kun Xu, and David J. Moss, "Photonic multiplexing techniques for neuromorphic computing", Nanophotonics vol. 12 (5): 795–817 (2023). DOI:10.1515/nanoph-2022-0485.

151. Chawaphon Prayoonyong, Andreas Boes, Xingyuan Xu, Mengxi Tan, Sai T. Chu, Brent E. Little, Roberto Morandotti, Arnan Mitchell, David J. Moss, and Bill Corcoran, "Frequency comb distillation for optical superchannel transmission", Journal of Lightwave Technology vol. 39 (23) 7383-7392 (2021). DOI: 10.1109/JLT.2021.3116614.

152. Mengxi Tan, Xingyuan Xu, Jiayang Wu, Bill Corcoran, Andreas Boes, Thach G. Nguyen, Sai T. Chu, Brent E. Little, Roberto Morandotti, Arnan Mitchell, and David J.

Moss, "Integral order photonic RF signal processors based on a soliton crystal micro-comb source", IOP Journal of Optics vol. 23 (11) 125701 (2021). https://doi.org/10.1088/2040-8986/ac2eab

153. Yang Sun, Jiayang Wu, Yang Li, Xingyuan Xu, Guanghui Ren, Mengxi Tan, Sai Tak Chu, Brent E. Little, Roberto Morandotti, Arnan Mitchell, and David J. Moss, "Optimizing the performance of microcomb based microwave photonic transversal signal processors", Journal of Lightwave Technology vol. 41 (23) pp 7223-7237 (2023). DOI: 10.1109/JLT.2023.3314526.

154. Mengxi Tan, Xingyuan Xu, Andreas Boes, Bill Corcoran, Thach G. Nguyen, Sai T. Chu, Brent E. Little, Roberto Morandotti, Jiayang Wu, Arnan Mitchell, and David J. Moss, "Photonic signal processor for real-time video image processing based on a Kerr microcomb", Communications Engineering vol. 2 94 (2023). DOI:10.1038/s44172-023-00135-7

155. Mengxi Tan, Xingyuan Xu, Jiayang Wu, Roberto Morandotti, Arnan Mitchell, and David J. Moss, "Photonic RF and microwave filters based on 49GHz and 200GHz Kerr microcombs", Optics Communications, vol. 465, Article: 125563 (2020). doi:10.1016/j.optcom.2020.125563. doi.org/10.1063/1.5136270.

156. Yang Sun, Jiayang Wu, Yang Li, Mengxi Tan, Xingyuan Xu, Sai Chu, Brent Little, Roberto Morandotti, Arnan Mitchell, and David J. Moss, "Quantifying the Accuracy of Microcomb-based Photonic RF Transversal Signal Processors", IEEE Journal of Selected Topics in Quantum Electronics vol. 29 no. 6, pp. 1-17, Art no. 7500317 (2023). 10.1109/JSTQE.2023.3266276.

157. Kues, M. et al. "Quantum optical microcombs", Nature Photonics vol. 13, (3) 170-179 (2019). doi:10.1038/s41566-019-0363-0

158. C.Reimer, L. Caspani, M. Clerici, et al., "Integrated frequency comb source of heralded single photons," Optics Express, vol. 22, no. 6, pp. 6535-6546, 2014.

159. C. Reimer, et al., "Cross-polarized photon-pair generation and bi-chromatically pumped optical parametric oscillation on a chip", Nature Communications, vol. 6, Article 8236, 2015.   DOI: 10.1038/ncomms9236.

160. L. Caspani, C. Reimer, M. Kues, et al., "Multifrequency sources of quantum correlated photon pairs on-chip: a path toward integrated Quantum Frequency Combs," Nanophotonics, vol. 5, no. 2, pp. 351-362, 2016.

161. C. Reimer et al., "Generation of multiphoton entangled quantum states by means of integrated frequency combs," Science, vol. 351, no. 6278, pp. 1176-1180, 2016.

162. M. Kues, et al., "On-chip generation of high-dimensional entangled quantum states and their coherent control", Nature, vol. 546, no. 7660, pp. 622-626, 2017.

163. P. Roztocki et al., "Practical system for the generation of pulsed quantum frequency combs," Optics Express, vol. 25, no. 16, pp. 18940-18949, 2017.

164. Y. Zhang, et al., "Induced photon correlations through superposition of two four-wave mixing processes in integrated cavities", Laser and Photonics Reviews, vol. 14, no. 7, pp. 2000128, 2020. DOI: 10.1002/lpor.202000128

165. C. Reimer, et al., "High-dimensional one-way quantum processing implemented on d-level cluster states", Nature Physics, vol. 15, no.2, pp. 148–153, 2019.

166. P.Roztocki et al., "Complex quantum state generation and coherent control based on integrated frequency combs", Journal of Lightwave Technology vol. 37 (2) 338-347 (2019).

167. S. Sciara et al., "Generation and Processing of Complex Photon States with Quantum Frequency Combs", IEEE Photonics Technology Letters vol. 31 (23) 1862-1865 (2019). DOI: 10.1109/LPT.2019.2944564.

168. Stefania Sciara, Piotr Roztocki, Bennet Fisher, Christian Reimer, Luis Romero Cortez, William J. Munro, David J. Moss, Alfonso C. Cino, Lucia Caspani, Michael Kues, J. Azana, and Roberto Morandotti, "Scalable and effective multilevel entangled photon states: A promising tool to boost quantum technologies", Nanophotonics vol. 10 (18), 4447–4465 (2021). DOI:10.1515/nanoph-2021-0510.

169. L. Caspani, C. Reimer, M. Kues, et al., "Multifrequency sources of quantum correlated photon pairs on-chip: a path toward integrated Quantum Frequency Combs," Nanophotonics, vol. 5, no. 2, pp. 351-362, 2016.